\documentclass[a4paper,fleqn]{cas-dc}

\usepackage{graphicx}
\usepackage{dcolumn}
\usepackage{bm}
\usepackage[english]{babel}
\usepackage{subfig}
\babeltags{en = english}
\usepackage{url}
\usepackage{tabularx}
\usepackage[square,comma,authoryear]{natbib}
\usepackage{placeins}
\usepackage{threeparttable}  
\usepackage{array} 
\usepackage{cancel}

\usepackage{xcolor}

\begin{document}

\title[mode=title]{A Scalable OpenLB–LAMMPS Framework for Fully-Resolved Simulations of Hindered Settling of Arbitrary Non-Spherical Particles}
\shorttitle{OpenLB-LAMMPS coupling}
\shortauthors{Babu et~al.}
\author[1]{Varghese Babu}
\affiliation[1]{
organization={Department of Materials Science and Chemical Engineering, State University of New York},
                addressline={100 Nicolls Road}, 
                city={Stony Brook},
                postcode={11794-2275}, 
                state={New York},
                country={USA}
}


\author[2]{Adrian Kummerländer}
\author[2]{Mathias J. Krause}
\affiliation[2]{
 organization={Lattice Boltzmann Research Group, Institute for Applied and Numerical Mathematics, Karlsruhe Institute of Technology},
 city = {Karlsruhe},
 country = {Germany}
}

\author[3]{Santosh Ansumali}
\affiliation[3]{
organization={Engineering Mechanics Unit,
Jawaharlal Nehru Centre for Advanced Scientific Research},
city={Jakkur, Bengaluru},
postcode={560064},
state = {Karnataka},
country={India}
}

\author[1]{Dilip Gersappe}

\date{\today}


\begin{abstract}
Hindered settling of non spherical particles remains significantly less understood than spherical particles due to the
computational challenges in resolving the complex contact mechanics and hydrodynamic interactions. In this paper, we present a scaleable fluid structure interaction (FSI) framework coupling the 
open-source LBM software in OpenLB \citep{olbPaper2021} with the Discrete Element Method (DEM) implemented in LAMMPS 
\citep{thompsonLAMMPSFlexibleSimulation2022} to simulate arbitrary 
shaped rigid bodies in a fluid. Particle contacts are captured using a multi-sphere ``clump'' representation in DEM, while particle geometries are resolved on the fluid grid via voxelization. We validate our model against single and multi-particle benchmarks, and proceed to study the hindered setttling of cubes in systems 
containing upto $\mathcal{O}(10^5)$ cubes. Our simulations show distinct differences between the settling of cubes and sphere, as cubes form pronounced coordination shells without face-parallel contact, in contrast to the contact-dominated clustering observed in spheres. Our results find that correlation length 
in velocity fluctuations scale with system size even for the largest 
system studied. These 
findings highlight the role of particle morphology in suspension dynamics and 
demonstrate a robust framework for simulating large-scale geotechnical and particulate flows. 
 
\end{abstract}

\maketitle


\section{Introduction}


Hindered 
settling is the phenomenon where particle-particle and particle-fluid interactions retard the downward velocity of suspensions with high particle concentrations. 
Found in many natural and industrial systems, this phenomenon has been subject to extensive 
experimental and numerical studies. 
However, numerical models of hindered settling has 
largely focused on spherical particles. This simplification 
is inherently limited, as it fails to capture the  contact 
dynamics, interlocking behaviors, and orientational microstructures.
In this 
work, we establish a 
computational framework to 
bridge this gap. We 
have combined  a fluid solver, OpenLB, with granular simulation 
package in LAMMPS, to accurately resolve arbitrary rigid-body 
shapes and their contact dynamics. 
This coupled framework for fluid-structure interaction
(FSI)  is first validated 
through the study of the hindered settling velocities of 
sedimenting cubes. 


Perhaps the best known result associated with hindered settling is 
the Richardson-Zaki relation which relates the bulk settling 
velocity of a suspension of spheres to the packing density \citep{richardsonSedimentationSuspensionUniform1954}. This relationship 
has been validated for spheres through experimental 
\citep{garside1977velocity,chongEffectParticleShape1979} and numerical evidence 
\citep{rettingerCoupledLatticeBoltzmann2017,willenContinuityWavesResolvedparticle2017,yinHinderedSettlingVelocity2007}. 
Owing to the numerical difficulties in contact detection, there are fewer studies on 
sedimentation of non-spherical shapes. Some of the recent studies on hindered settling of non-spherical shapes \citep{seyed-ahmadiSedimentationInertialMonodisperse2021,kunduSettlingDynamicsNonBrownian2025,zhangLatticeBoltzmannSimulations2016, fareedCollectiveSedimentationSymmetric2026}. 

Simulations of 
particles in a fluid typically fall into Euler-Euler or Euler-Lagrange 
methods. Euler-Euler methods represent the particle phase as continuum.  In Euler-Lagrange methods, particles are evolved individually. 
When particles are simulated individually, the predominant 
method for simulating particles is DEM 
\citep{cundallDiscreteNumericalModel1979}. These simulations 
are often called CFD-DEM.

There are several commercial and open-source implementation that 
are available for CFD-DEM. 
In  recent years, 
Lattice-Boltzmann methods (LBM) have become increasingly prominent 
and have emerged as a alternative to solving the Navier-Stokes equation 
directly. LBM has the advantages of being more easily parallelizable 
and being well-suited for simulating complex geometries.  
There are different methods developed for particle-fluid interaction 
in a LBM simulation. They largely fall into momentum-exchange methods (MEM)
\citep{laddNumericalSimulationsParticulate1994,laddNumericalSimulationsParticulate1994a} and partially saturated methods (PSM) \citep{nobleLatticeBoltzmannMethodPartially1998}. 
In MEM, a no-slip boundary condition is applied along the surface 
of the particle. This is in contrast with PSM where each node in 
the lattice is parametrized by solid fraction and the collision 
operator in LBM is modified according to the solid fraction. See 
\citep{rettingerComparativeStudyFluidparticle2017,trunkRevisitingHomogenizedLattice2021} for a comparison 
between the methods. Various implementations of CFD-DEM are listed in Table \ref{tab:cfd_dem_summary}.
\begin{table*}[htbp]
\centering
\small
\begin{tabularx}{\textwidth}{L L L l}
\toprule
\textbf{Framework / Software} & \textbf{Fluid Solver} & \textbf{DEM Engine} & \textbf{License Type} \\
\midrule
\multicolumn{4}{l}{\textbf{Direct Navier-Stokes (CFD-DEM) Implementations and LBM-DEM implementations}} \\
\midrule
\makecell{CFDDEM-Coupling \\ \citep{goniva2012influence} } & OpenFOAM \newline \citet{weller1998tensorial} & LIGGGHTS \newline \citep{kloss2012models} & Open-Source \\ \addlinespace
\makecell{YADE \\ \citep{smilauer2023yade}} & OpenFOAM  & YADE (Built-in) & Open-Source \\ \addlinespace
\makecell{MFiX \\ \citep{syamlal1993mfix}} & Built-in & Built-in & Open-Source \\ \addlinespace
\makecell{Lethe-DEM \\ \citep{blaisLetheOpensourceParallel2020}} & Built-in & Built-in & Open-Source \\ \addlinespace
\makecell{openHFDIB-DEM \\ \citep{studenikOpenHFDIBDEMExtensionOpenFOAM2024}} & OpenFOAM & Built-in & Open-Source \\ \addlinespace
\makecell{Ansys Fluent + Rocky \\ \citep{fluent2023ansys}} & Built-in & Ansys Rocky & Commercial \\ \addlinespace
\makecell{Commercial CFDDEM-coupling} & OpenFOAM & LIGGGHTS & Commercial \\ \addlinespace
\makecell{LBDEMcoupling \\ \citep{seil2016lbdemcoupling}} & LBM (Palabos \citep{lattPalabosParallelLattice2021}) & LIGGGHTS & Open-Source \\ \addlinespace
\makecell{LAMMPS-LBM-PSM \\ \citep{najuchAnalysisTwoPartiallysaturatedcell2019a}} & LBM (Integrated) & LAMMPS  & Open-Source \\ \addlinespace
\makecell{LIGGGHTS-Palabos \\ \citep{ahmadianSimulatingFluidSolid2024}} & LBM (Palabos) & LIGGGHTS & Open-Source  \\ \addlinespace
\makecell{OpenLB \\ \citep{krauseParticleFlowSimulations2017}} & LBM (OpenLB) & Built-in  & Open-Source   \\ \addlinespace
\makecell{LEDDS \\ \citep{maggio-aprileLEDDSPortableLBMDEM2025}} & LBM  & Built-in & Open-Source \\ \addlinespace
\makecell{WaLBerla \\ \citep{bauerWaLBerlaBlockstructuredHighperformance2021}}& LBM (Block-structured) & Built-in & Open-Source\\ \addlinespace
\bottomrule
\end{tabularx}
\caption{Summary of Commercial and Open-Source CFD-DEM}
\label{tab:cfd_dem_summary}
\end{table*}

Many of the implementations in Table \ref{tab:cfd_dem_summary} are capable of simulating 
non-spherical rigid bodies. The main challenge in simulating 
irregular shaped rigid bodies is contact detection. There are various methods developed for 
contact detection between general non-spherical shapes. The most 
common methods are multi-sphere (clump) representation, 
superquadratics, polyhedral methods and using spherical harmonics. 
Each method comes with its own advantages and disadvantages, however the ``clump'' method is the easiest and the most straightforward method to implement. 
It is also very flexible in that any shape can be decomposed into a collection of spheres. The major drawback of this method is that one might need a large number of spheres for an accurate representation.
See \citep{zhaoRevolutionizingGranularMatter2023, 
capozzaHierarchicalSphericalHarmonicbased2021, 
luDiscreteElementModels2015} for more details. 

Despite various available implementations, simulating $\mathcal{O}
(10^5)$ arbitrary rigid bodies still remains  a major  computational challenge. To the best of authors knowledge,  openHFDIB-DEM
\citep{studenikOpenHFDIBDEMExtensionOpenFOAM2024} is the only framework that has demonstrated 
such a capability. To address this gap, in this article we couple LAMMPS 
\citep{thompsonLAMMPSFlexibleSimulation2022} and OpenLB 
\citep{olbPaper2021}, both designed and capable of large 
scale parallel simulations. We use the clump multi-sphere 
representation in LAMMPS for rigid bodies and make use of the FSI 
(fluid-structure interaction) module of OpenLB to create a fully-
resolved LBM-DEM simulation of arbitrary shaped rigid bodies. 
Presence of the arbitrary solid bodies
is represented as a time-dependent lattice porosity field
  on the Eulerian grid.

We use the LAMMPS-OpenLB framework to study hindered settling of cubes 
with the aim of understanding the microstructure. We first validate
our framework through simulations of spheres and highlight the differences between cubes and spheres during sedimentation. 


\section{Models and Methods}
\begin{figure}[htbp]
    \centering
    \includegraphics[width=\linewidth]{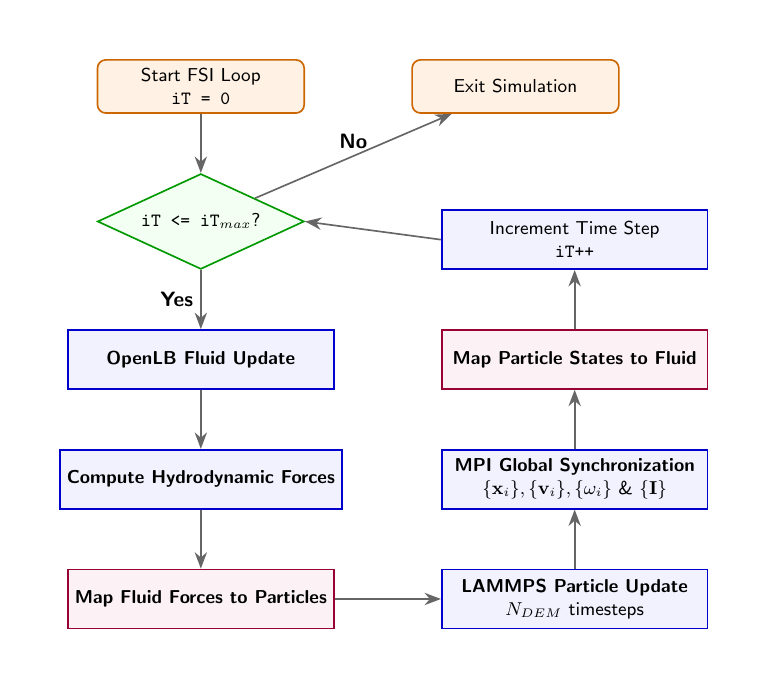}
    \caption{\textbf{Flow chart describing our implementation of OpenLB-LAMMPS coupling}. $\{{\bf{x}}_i\},\{{\bf{v}}_i\},\{{\bf{\omega}}_i\}$ and $\{{\bf{I}}_i\}$ are positions, velocities, angular momentums and the rotated inertia matrix respectively.}
    \label{fig:flowchart}
\end{figure}

The simulation consists of using the Lattice-Boltzmann Method (LBM) 
for the fluid and Discrete-Element-Method (DEM) for the rigid 
bodies. An important feature of these simulations is the ``clump'' 
representation of rigid bodies. We use the ``CLUMP'' package to 
generate the overlapping sphere representation of any arbitrary shape \citep{angelidakisCLUMPCodeLibrary2021}. The rigid 
bodies are represented as a porosity field in LBM using the FSI module and therefore, 
we can use the irregular shaped rigid body (available to us as an 
.stl file, for example) directly in LBM. This means that the 
effect of any particular clump representation of rigid bodies is 
only indirectly felt by fluid. This crucially differentiates our 
work from similar work in \citep{ahmadianSimulatingFluidSolid2024} where the rigid body is represented as a collection of spheres in the fluid as well. 
Representation of the rigid body as sphere clump in the fluid 
introduces an additional approximation as the fluid-structure 
interactions are influenced by the clump representation. 




\subsection{OpenLB-LAMMPS coupling}

The simulation consists of running LBM and DEM simulations 
sequentially. Depending on the details of the particle contact, 
viscosity of the fluid and the chosen relaxation constant $\tau$, 
we have two values of timesteps in the simulation $\Delta 
t_{\mathrm{DEM}}$ and $\Delta t_{\mathrm{LBM}}$. For most of the cases we have 
considered here, $\Delta t_{\mathrm{DEM}} < \Delta t_{\mathrm{LBM}}$. Therefore, we 
simulate $n_{\mathrm{DEM}}=\frac{\Delta t_{\mathrm{LBM}}}{\Delta t_{\mathrm{DEM}}}$ DEM 
timesteps for every LBM timestep. During these $n_{\mathrm{DEM}}$ 
timesteps, the forces on the DEM particle due to the fluid is 
kept constant. Forces on the fluid are calculated based on the final positions and velocities of the particles. 
In the other case where $\Delta t_{\mathrm{DEM}} > \Delta 
t_{\mathrm{LBM}}$, we simply set $\Delta t_{\mathrm{DEM}} = \Delta t_{\mathrm{LBM}}$. 

The full simulation involves exchange of information between the 
fluid and the rigid bodies. The fluid solver calculates 
hydrodynamic forces/torques acting on the rigid bodies and 
these are included in DEM simulation as external forces/torques. 
After the DEM simulations are completed, the new positions, 
velocities, orientations and angular velocities are communicated 
back to the fluid simulation. This is described in the flowchart Fig. \ref{fig:flowchart}.

We use OpenLB \citep{olbPaper2021} for LBM and 
GRANULAR package in LAMMPS \citep{thompsonLAMMPSFlexibleSimulation2022} for 
DEM simulations. LAMMPS is used as 
a library with minimal modifications to its source code. Note 
that although the rigid body is represented in LAMMPS as a 
collection of overlapping spheres, this representation is only used by LAMMPS 
for contact detection and force calculation. The dynamics of the 
rigid body due to contact forces (and the external forces from 
LBM) is determined by the mass and moments of inertia specified 
independently through  the \texttt{infile} feature of rigid body 
simulation in LAMMPS. In all the simulations reported here, we 
use the BGK collision operator for the fluid phase and 
Hertzian interactions with tangential friction for the 
particle contacts.

For the fluid side of the FSI, we use a homogenized lattice Boltzmann method (HLBM) based on the Brinkman-Navier-Stokes equations (BNSE) \citep{krauseParticleFlowSimulations2017,trunkRevisitingHomogenizedLattice2021}. The bodies are represented on a fixed Eulerian lattice by a time-dependent porosity field $d(\bm{x},t)$, and the no-slip condition is imposed through a volume penalization term (Brinkman drag) added to the incompressible Navier-Stokes equations. For a fluid with kinematic viscosity $\nu$, density $\rho$, and pressure $p$, this monolithic formulation is governed by:
\begin{equation}
    \frac{\partial \bm{u}}{\partial t} + \bm{u} \cdot \bm{\nabla} \bm{u} = -\frac{\bm{\nabla} p}{\rho} + \nu \bm{\nabla}^2 \bm{u} - \frac{\nu}{K(\bm{x}, t)}(\bm{u} - \bm{u}_{\mathrm{s}})
\end{equation}
where $\bm{u}$ is the fluid velocity and $\bm{u}_{\mathrm{s}}$ is the local solid velocity. The permeability $K(\bm{x},t)$ is a function of the porosity field, such that the penalization term vanishes in the fluid, where $K \to \infty$, and drives $\bm{u}$ towards $\bm{u}_{\mathrm{s}}$ inside the bodies, where $K$ takes its lowest numerically stable value. The motion of the bodies, changes of their geometrical arrangement and their contacts enter the fluid solver only through $d(\bm{x},t)$, so that arbitrarily shaped bodies require no special treatment. The penalization is realized as a Kupershtokh forcing term in the collision step, while the hydrodynamic forces and torques acting on the bodies are obtained from a momentum exchange over the lattice links crossing the fluid-solid interface.
This Brinkman penalization approach has recently been successfully applied to complex FSI problems, such as wall-modeled large eddy simulations of wind turbine rotors \citep{kummerlaenderHFF2026} and vocal fold oscillations \citep{kummerlaenderCMAME2026}.

During the coupled simulation, the structural geometry is resolved via an exact surface voxelization in the fluid solver, while LAMMPS concurrently tracks the contact mechanics and trajectories using the multi-sphere clump representation.
This separation allows us to handle complex, arbitrarily shaped geometries contacts.
This distinguishes our approach from prior HLBM implementations in OpenLB, such as the work by Trunk et al.~\citep{trunkRevisitingHomogenizedLattice2021}, which lacked a collision model, and Marquardt et al.~\citep{marquardtNovelModelDirect2024a}, which was limited to convex bodies.
The scheme is implemented in the open source LBM framework OpenLB \citep{olbPaper2021} as platform-transparent operators, so that the same model definition is executed on CPUs and GPUs. LBM and DEM simulations are detailed in Appendix \ref{MethodsDetails}.
\section{Validation}

\begin{figure}[htbp]
\centering
\includegraphics[width=0.9\linewidth]{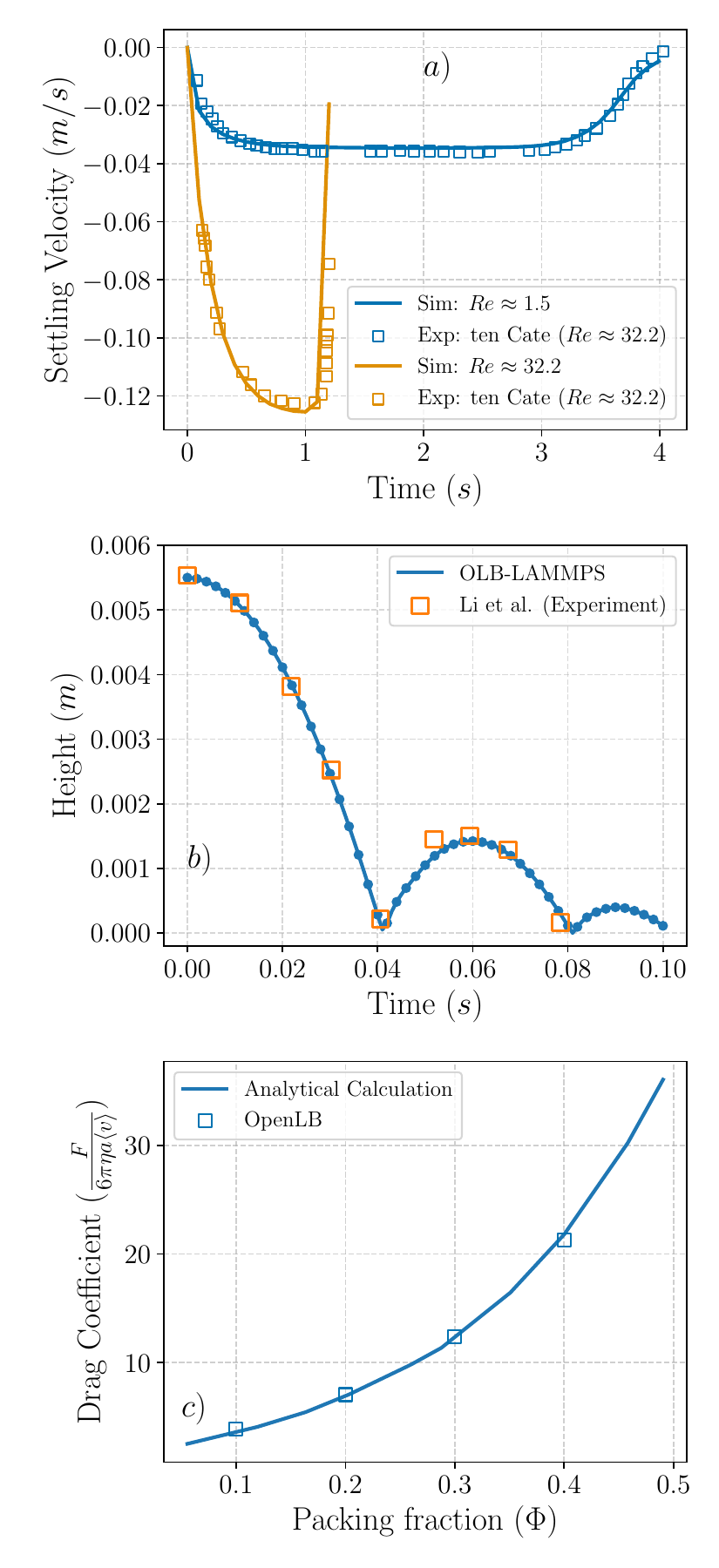}
\caption{\textbf{Sphere simulation in a fluid.} \textbf{a)} Comparison of experimental \citep{tencateParticleImagingVelocimetry2002} and simulated settling velocities for a sphere falling in a viscous liquid. \textbf{b)} Comparison of experimental results with numerical results of a sphere-wall collision in viscous liquid.\textbf{c)} Comparison of the drag coefficient calculated from analytical calculation and from numerical simulations for  a periodic array of spheres.}
\label{fig:ValidationSpheres}
\end{figure}

\begin{table}[htbp]
\centering
\footnotesize
\begin{threeparttable}
\begin{tabular}{
  >{\raggedright\arraybackslash}p{3.3cm}
  >{\centering\arraybackslash}p{1.1cm}
  >{\raggedright\arraybackslash}p{2.6cm}
}
\hline
Parameter & Symbol & Value \\
\hline
\multicolumn{3}{l}{\textbf{Terminal velocity of single sphere}} \\
Sphere radius & $R$ & $7.5\times10^{-3}$ m \\
Sphere density & $\rho_s$ & $1120$ kg/m$^3$ \\
Fluid density (E1/E2) & $\rho_f$ & $970$/$960$ kg/m$^3$ \\
Fluid viscosity (E1/E2) & $\mu$ & $0.373$/$0.058$ Pa$\cdot$s \\
Reynolds number (E1/E2) & $Re$ & $\approx1.5$/$\approx32.2$ \\
Relaxation time (E1/E2) & $\tau$ & 0.661/0.6 \\
Resolution of $R$  & $N_{\mathrm{res}}$ & 9 \\
\hline
\multicolumn{3}{l}{\textbf{Sphere-wall collision in fluid}} \\
Sphere diameter & $d_p$ & $9.5\times10^{-3}$ m\tnote{1} \\
Sphere density & $\rho_{\mathrm{sphere}}$ & $7780$ kg/m$^3$ \\
Fluid density & $\rho_{\mathrm{fluid}}$ & $1203$ kg/m$^3$ \\
Fluid viscosity & $\mu$ & $50.2\times10^{-3}$ Pa$\cdot$s \\
Young's modulus (S/W) & $E$ & $20$/$3.3$ GPa \\
Poisson's ratio (S/W) & $\nu$ & $0.33$/$0.24$ \\
Domain dimensions & $L_x{\times}L_y{\times}L_z$ & $42{\times}42{\times}50$ mm$^3$ \\
Sphere resolution & $d_p/\Delta x$ & $32$ l.u. \\
\hline
\multicolumn{3}{l}{\textbf{Drag force, periodic array}} \\
Domain size & $L$ & $1$ m \\
Grid resolution/dim. & $N$ & $64$ \\
Viscosity & $\mu$ & $0.02$ Pa$\cdot$s \\
LBM relaxation time & $\tau$ & $0.55$ \\
\hline
\multicolumn{3}{l}{\textbf{Settling velocity of cube}} \\
Fluid density & $\rho_f$ & $1135.19$ kg/m$^3$ \\
Kinematic viscosity & $\nu$ & $7.325{\times}10^{-6}$ m$^2$/s \\
Particle density & $\rho_p$ & $1221.0$ kg/m$^3$ \\
Cube length & $a$ & $0.79$, $1.00$, $1.20$, $1.36$, $1.59$ mm \\
Domain size & $L_x,L_y,L_z$ & $20a{\times}20a{\times}50a$ \\
LBM relaxation time & $\tau$ & $1.0$ \\
Resolution & $a/\Delta x$ & $16$ \\
\hline
\end{tabular}
\end{threeparttable}
\caption{Summary of physical and numerical parameters used across the three validation cases.}
\label{tab:all_sim_parameters}
\end{table}

We test our implementation by validating sphere settling 
simulation in a fluid. 
Our initial set of simulations model  three cases: the settling of a single sphere in a fluid, sphere-wall collision, and the calculation of the drag force on a periodic array of spheres. 

\subsection{Terminal velocity of a single sphere}
\label{tenCate}
We reproduce here the validation of the simplest case of a sphere 
settling in a fluid as discussed in 
\citet{tencateParticleImagingVelocimetry2002}. The experimental 
setup involves dropping a sphere in a viscous fluid. We compare 
``Case E 1'' and ``Case E 2'' in 
\citet{tencateParticleImagingVelocimetry2002} corresponding to a 
Reynolds number of $\approx 1.5$ and $\approx 32.2$ respectively. 
Our results as shown in Fig. \ref{fig:ValidationSpheres}a show a 
good agreement with experimental results. The parameters used for this simulation is presented in Table. \ref{tab:all_sim_parameters}.

\subsection{Sphere-wall collision in a fluid}
\label{collision}
We compare an experimental result of a metallic sphere bouncing 
on a surface in a viscous liquid from 
\citet{liContactModelNormal2012} 
with our numerical 
method in Fig. \ref{fig:ValidationSpheres} b. 
The presence of lubrication force as the sphere 
approaches the wall is not captured in our simulations as the 
gap between the two surfaces become smaller than the lattice 
separation. To match 
with the experimental data we enlarge the particle radius by 
$5\%$ as mentioned in \citet{marquardtDiscreteContactModel2023}. The result, which correspond to Case 1 of 
\citet{liContactModelNormal2012} 
is shown in Fig. 
\ref{fig:ValidationSpheres} b. As the sphere comes close to the wall lubrication forces become important - we discuss the lubrication force in Appendix \ref{LubricationForce}.

\subsection{Drag force on a periodic array of spheres.}
\label{ladd1}
Another fundamental result relevant for validation of our 
numerical method is the drag force on a periodic array of 
spheres. Analytical result exists for this calculation 
\citep{sanganiSlowFlowPeriodic1982,laddHydrodynamicInteractionsSuspension1988,hasimoto1959periodic}.
In this test, we follow the numerical setup 
outlined in \cite{rettingerCoupledLatticeBoltzmann2017}. We 
define a domain of $L\times L\times L$ with periodic boundary 
conditions and place a sphere at the center whose radius is 
determined by the packing fraction $\phi$ we are studying. We apply a 
constant force density to all the fluid nodes $f_{\mathrm{fluid}}$.
We then 
measure the force acting on the sphere and 
the average velocity in the domain. At steady state, the 
measured force on the sphere is equal to the total force 
applied on all fluid nodes. As discussed in 
\cite{rettingerCoupledLatticeBoltzmann2017}, in a real system 
where one measures the drag force on the sphere, that value will include the buoyancy force as well. 
However, since we are only 
applying the force to the fluid nodes, the drag force we 
measure in the sphere is due to the flow of the fluid alone. We 
need to add back the buoyancy force to the drag force to get the 
correct measurements. Therefore, total force is  $\boldsymbol{F}_t=\frac{\boldsymbol{F}_d}{1-\phi}$. Our 
results are shown in Fig. \ref{fig:ValidationSpheres}c. 
In Appendix \ref{DragForceLadd} we discuss the effect of 
$\tau$ and resolution on the results.

\subsection{Settling velocity of cuboid}

\label{settlingVelocityCuboid}
\begin{figure}
    \centering
    \includegraphics[width=\linewidth]{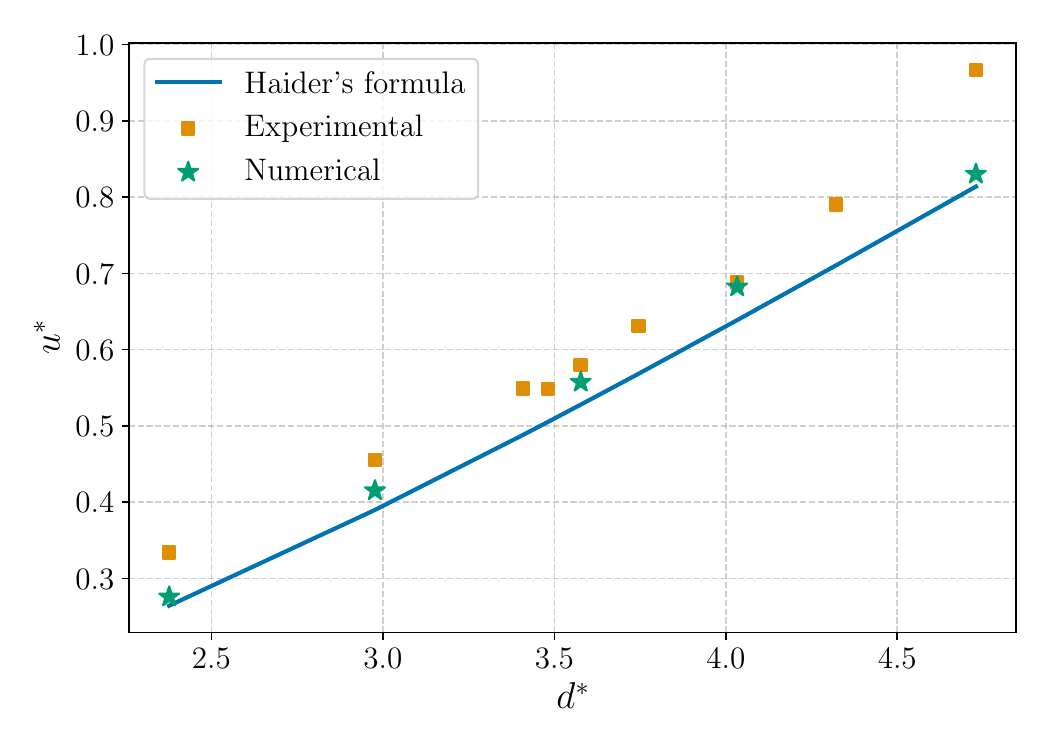}
    \caption{\textbf{Settling velocity of cubes}. We compare the values calculated from numerical method and empirical formula with experimental results from \cite{kwonExperimentalInvestigationSettling2025}.}
    \label{fig:cubeSettling}
\end{figure}
For validation of our code for non-spherical shapes, we compare our numerical technique with experimentally 
observed settling velocity of cuboidal particles as reported in 
\cite{kwonExperimentalInvestigationSettling2025}. The settling velocity of a cube can also be calculated using the empirical relation by \cite{haiderDragCoefficientTerminal1989}. 
This relationship is given by 
\begin{equation}
\label{haiderFormula}
    u_* = \left[ \frac{18}{d_*^2} + \frac{(2.3348 - 1.7439 \Phi)}{d_*^{0.5}} \right]^{-1}
\end{equation}
where $$u_*=u_p \left[\frac{\rho_f^2}{g \mu (\rho_s - \rho_f)} 
\right]^{1/3}$$ and $$d_* = d_{\mathrm{\text{sph}}} \left[ \frac{g \rho_f 
(\rho_s - \rho_f)}{\mu^2} \right]^{1/3}$$ with $u_p$ being the 
particle velocity, $d_{\mathrm{\text{sph}} }$ the equivalent spherical 
diameter of the cuboid and $\Phi$ the sphericity. The fluid 
and particle parameters are taken from the experimental study  
\cite{kwonExperimentalInvestigationSettling2025} and are given in 
Table \ref{tab:all_sim_parameters}. We consider the glycerol-water mixture studies for comparison. In the simulation we use pressure boundary conditions in
the $x$ and $y$ directions and we use bounce-back boundary condition on $z$ walls. 
As shown in Fig. \ref{fig:cubeSettling}, the numerical 
method underestimates settling velocity when compared to the experimental result. 
We also show the empirical relationship alongside, which is lower than both numerical 
and experimental results 
As noted in \cite{kwonExperimentalInvestigationSettling2025}, 
the cubes maintain their initial orientation throughout the decent. 

\section{Results and Discussion}

\subsection{Hindered settling of Cubes and Spheres}
\label{sec:hinderedSettling}
\begin{table}
\begin{tabular}{lcr}
\textbf{Parameter} & \textbf{Symbol} & \textbf{Value} \\
Number of spheres  & $N_{\mathrm{spheres}}$ & 3150  \\
Fluid density & $\rho_f$ & $1000$ kg/m$^3$ \\
Fluid viscosity & $\mu$ & $10^{-3}$ Pa$\cdot$s \\
Particle density & $\rho_p$ & $2500$ kg/m$^3$ \\
Sphere diameter & $d_{\mathrm{sphere}}$ & $0.35 \times 10^{-3}$ m \\
Cube dimension & $a$ & $0.283 \times 10^{-3}$ m \\
Young's modulus & $E$ & $5\times10^{4}$ Pa \\
Poisson ratio & $\nu$ & $0.3$ \\
Restitution coefficient & $e$ & $0.9$ \\
Friction coefficient  & $\mu_t$ & $0.6$ \\
Reynolds Number & $Re$ & 16.8 \\
Resolution of $d_{\mathrm{sphere}}$ & $N_{\mathrm{res}}$ & 12 \\
LAMMPS time step & $\Delta t$ & $0.05 t_c$ \\
LBM relaxation time & $\tau$ & 0.55 \\
\hline
\end{tabular}
\caption{\label{tab:hinderedSettlingParameters} Summary of physical and numerical parameters used for hindered settling of cubes and spheres. 
$G = \frac{E}{2(1 + \nu)}$ and $\beta = 0.8766 + 0.1630\nu$. LAMMPS 
contact time $t_c$ is derived from the material properties and particle size: $\pi \frac{d_{\mathrm{sphere}}}{2\beta}\sqrt{\frac{\rho_p}{G}}$ known as Rayleigh Time Step.}
\end{table}
In this section we present the results from the numerical study of hindered settling of spheres 
and cubes. For comparison between spheres and cubes, we use cubes that 
are volume-equivalent to spheres while keeping all other parameters fixed. 
Settling velocity of a collection of spheres at different packing 
fractions has been studied numerically 
\cite{rettingerCoupledLatticeBoltzmann2017,marquardtNovelModelDirect2024a,marquardtReviewHomogenizedLattice2024,trunkRevisitingHomogenizedLattice2021,yinHinderedSettlingVelocity2007}. 
Settling dynamics of multiple cubes was studied numerically in 
\cite{kunduSettlingDynamicsNonBrownian2025,marquardtNovelModelDirect2024a,marquardtNovelParticleDecomposition2024,seyed-ahmadiSedimentationInertialMonodisperse2021,fareedCollectiveSedimentationSymmetric2026} experimentally in \cite{paulInfluenceShapeSurface2017}. 

The simulation is conducted in a domain that is 
periodic in three directions. To prevent the particles from accelerating 
infinitely due to gravity, we apply a force to the fluid.
This force is periodically adjusted such that the total velocity of 
the fluid is zero. We use the following equation to update the 
compensating force  
$$f^{new}_{\mathrm{\alpha}} = f^{old}_{\mathrm{\alpha}} - K_p \langle v_{\mathrm{\alpha}} \rangle$$ with $K_p = 10^{-2}$ and $\alpha \in [x,y,z]$. This update 
is performed every $10^{-4}$s. The complete set of simulation parameters is specified in 
Table \ref{tab:hinderedSettlingParameters}. 

This numerical method of studying hindered settling using triply-periodic
box with a compensating force has been used in various 
previous studies such as 
\citet{laddHydrodynamicScreeningSedimenting1996,uhlmannSedimentationDiluteSuspension2014,rettingerCoupledLatticeBoltzmann2017,fareedCollectiveSedimentationSymmetric2026,fornariSedimentationFinitesizeSpheres2016,yinHinderedSettlingVelocity2007}. 

To generate a non-overlapping initial packing, we ran a long 
simulation (timesteps$>10^{7}$ or $\sim 220$ turnover times) starting from a configuration where spheres are arranged in a 
square lattice at $\phi=0.4$. 
The configuration generated is assumed to have randomized positions and 
are rescaled 
down to different packing fractions we study. The same initial condition is used for both spheres and cubes as the cubes are volume equivalent to the spheres.  
The hindered settling simulation requires a brief period for the compensating force to stabilize, after which the average velocities  of the system remain constant. We notice that the system is characterized by a small 
drift in the perpendicular direction to gravity which is of an order of magnitude 
smaller than the bulk terminal velocity. Once the $\langle v_{\mathrm{z}} \rangle$ values have reached their steady state values, we use the 
last 100 configurations from the trajectory for our analysis. 
The clump representation of the cube we study and a validation of it is given in Appendix \ref{validclumosection}.

\subsubsection{Hindered Settling velocity}
\begin{figure}
    \centering
    \includegraphics[width=\linewidth]{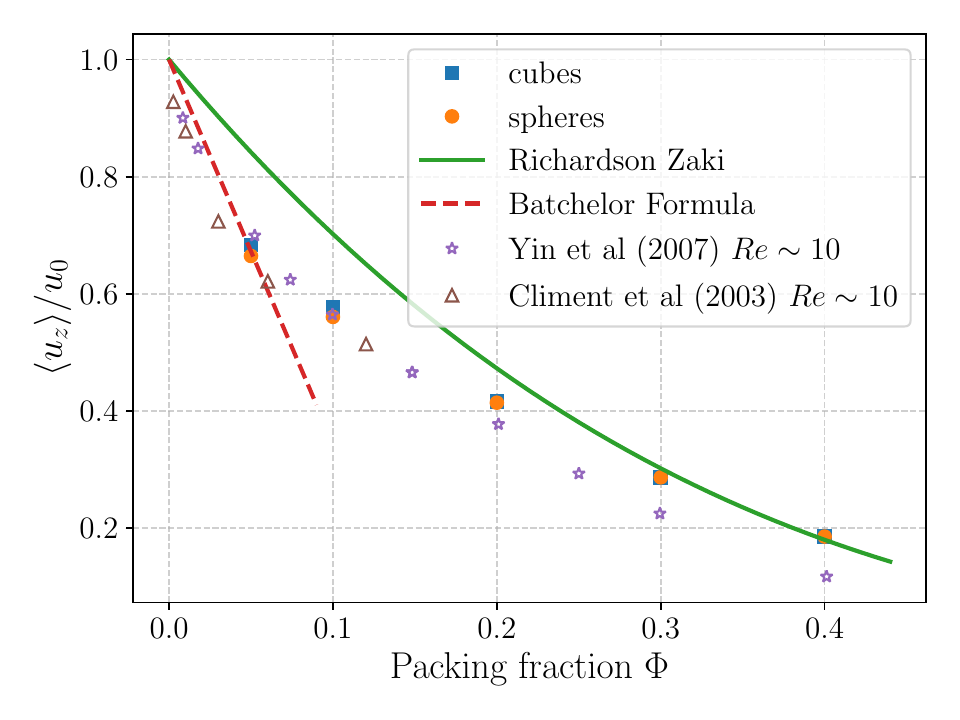}
    \caption{\textbf{Hindered settling velocities of cubes and spheres at different packing fractions.} We also show comparison with \citet{yinHinderedSettlingVelocity2007,climentNumericalSimulationsRandom2003}} $u_0$ for spheres is $-0.0488$m/s and $-0.0457$m/s for cubes.
    \label{fig:hinderedSettlingCubesSpheres}
\end{figure}

\label{hinderedsettlingspheres}

The dependence of the settling velocity of a collection of spheres on the packing density $\phi$ is empirically given by Richardson-Zaki relation \citep{richardsonSedimentationSuspensionUniform1954}. 
The relation is 
\begin{equation}
    \frac{\langle u_p \rangle}{u^*} = (1-\phi)^{\kappa}
    \label{eq:richardsonZaki}
\end{equation}
where 
\[
\kappa=\begin{cases}
4.65 & Re<0.2\\
4.35Re^{-0.03} & 0.2<Re<1\\
4.45Re^{-0.1} & 1<Re<500\\
2.39 & 500\leq Re
\end{cases}
\]. Here $\langle u_p \rangle$ is the particle velocity and $u^{*}$ is the settling velocity of a single particle. First we calculate the terminal velocity of a single cube and a single 
sphere as $\sim 0.0457$m/s and $\sim 0.048$m/s respectively
(see Appendix \ref{spheretermianlVelocity},\ref{cubetermianlVelocity}). The terminal 
velocity for sphere agrees with numerical calculation in \citet{rettingerCoupledLatticeBoltzmann2017} for the same system parameters.

The settling velocities of cubes and spheres at different packing fraction is shown in Fig. \ref{fig:hinderedSettlingCubesSpheres}.
We also show the Richardson-Zaki relation calculated using $Re=16.8$. 
We also show the result from  \cite{batchelorSedimentationDiluteDispersion1972} 
valid for dilute suspensions. We also show results at similar $Re$ from \citet{climentNumericalSimulationsRandom2003} and \citet{yinHinderedSettlingVelocity2007}.
Numerical results diverge from the empirical formula 
especially at lower packing fractions. At higher packing 
fractions, our simulations predict a higher velocity than is 
found in previous studies of comparable $Re$. \citet{climentNumericalSimulationsRandom2003} and \citet{yinHinderedSettlingVelocity2007} both explicitly add 
lubrication forces to the study which could be why our results 
differ from previous at higher packing fractions.

\subsubsection{Radial Distribution Function (RDF)}
\begin{figure}
    \centering
    \includegraphics[width=\linewidth]{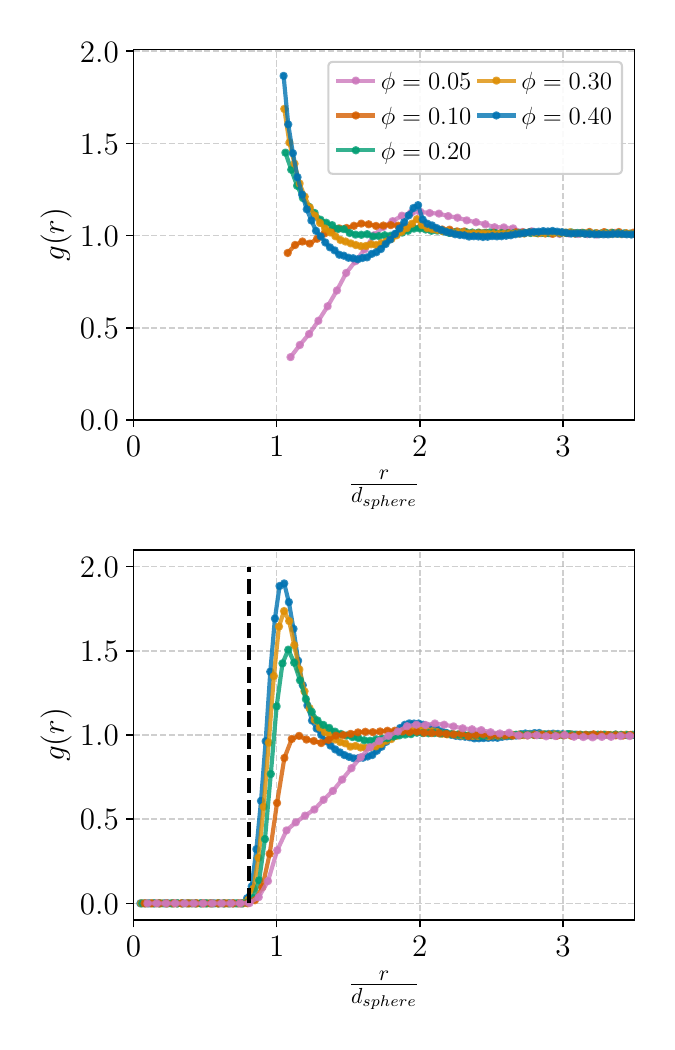}
    \caption{\textbf{Radial distributions of spheres (top) and cubes (bottom) for different packing fractions}. The black dashed line of the bottom figure indicates the side length of cube $a=0.000283$m.}
    \label{fig:corr_g_r}
\end{figure}

\begin{figure}
    \centering
    \includegraphics[width=\linewidth]{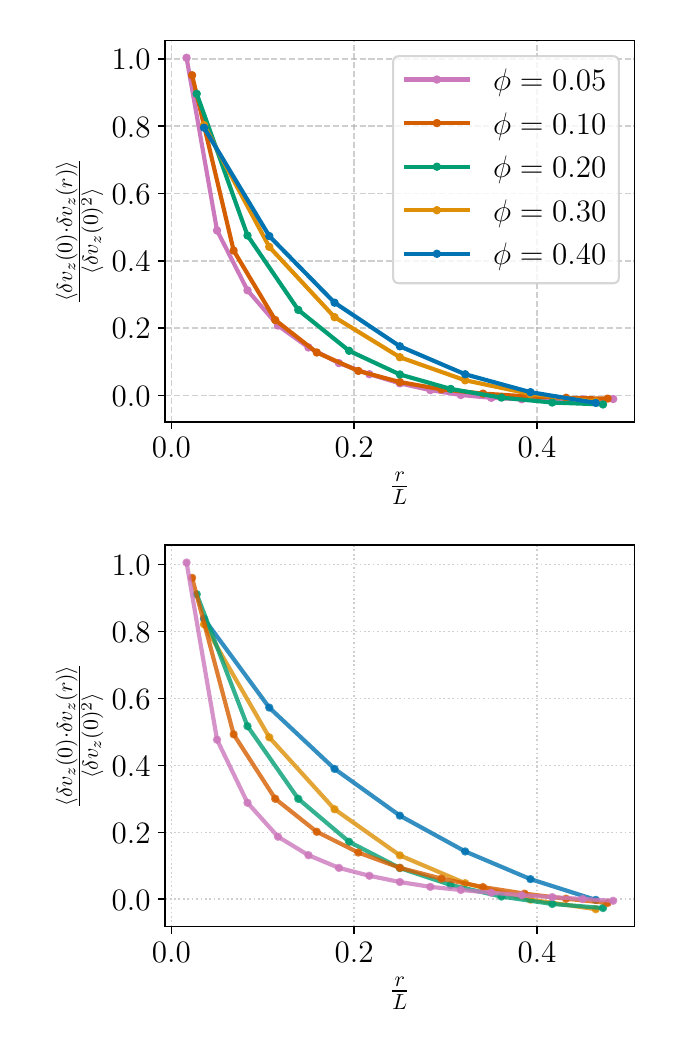}
    \caption{\textbf{Correlations in velocity fluctuations of the $z$ component for spheres (top) and cubes (bottom) for different packing fractions.} }
    \label{fig:corr_v}
\end{figure}

To understand the structure of cubes and spheres undergoing sedimentation we compute the radial distribution function as given by 
\begin{equation}
  g(r)=\frac{\langle N_{\mathrm{pairs}}(r,r+dr)\rangle}{\frac{N(N-1)}{2}\times\frac{4\pi r^{2}dr}{V}}
\end{equation} where the averaging is over configurations at 
steady state. The denominator term in the above expression is the 
number of pairs that will be found in a spherical shell of radius 
$r$ if the particle positions are generated from a uniform distribution. 
The results for cubes and spheres are shown in Fig. \ref{fig:corr_g_r} and Fig. \ref{fig:corr_v}. 
The RDFs clearly show the 
difference in structure formed by cubes and spheres. RDF of cubes clearly shows the formation of a first and a second shell. In addition, 
it is clear that the cubes are not touching each other with their 
faces parallel - then one would see the first peak before scaled 
distance $1$. For spheres, we observe that, except at the lowest 
packing fraction we studied, spheres tend to touch each other, with 
the number of pairs in contact increasing as packing fraction is 
increased. These figures can be directly compared with results in \citet{seyed-ahmadiSedimentationInertialMonodisperse2021,yinHinderedSettlingVelocity2007,kunduSettlingDynamicsNonBrownian2025,fareedCollectiveSedimentationSymmetric2026}. 
At low packing fractions $\phi<0.1$, RDF results of \citet{seyed-ahmadiSedimentationInertialMonodisperse2021} 
and \citet{kunduSettlingDynamicsNonBrownian2025} shows spheres in 
contact, which \citet{yinHinderedSettlingVelocity2007}'s RDF for $Re \sim 10$ and our results do not observe. Our RDF of cubes agrees 
quantitatively with \citet{seyed-ahmadiSedimentationInertialMonodisperse2021}'s result for $\phi=0.2$, 
while at lower packing fractions, their results predict much larger 
number near pair contacts. See comparison in Appendix \ref{comparisonAhmadiSection}. It must be noted that 
simulation in \citet{seyed-ahmadiSedimentationInertialMonodisperse2021} are at a higher 
Reynolds number than here. 


\subsubsection{2-point spatial correlation of velocity}
In the steady state, the other natural quantity to calculate is 
the velocity-velocity correlation in space. Since the system as a whole has 
a mean velocity in the $-\hat{z}$ direction due to gravity, we measure 
the correlation between velocity fluctuations. The fluctuation is 
calculated as $\delta v_z = v_z - v_{\mathrm{mean}}$ where $v_{\mathrm{mean}} = 
\langle v_z \rangle$. The correlation is calculated by 
$$C(r) = \frac{\langle \delta v_z(0) \cdot \delta v_z(r) \rangle}{\langle \delta v_z(0)^2 \rangle}.$$ 

We show the 2 point correlation of velocity in Fig. 
\ref{fig:corr_v}. Given the periodic 
boundary 
conditions, we calculate the correlation function only up-to $0.5 L$ where $L$ is the side length of the cubical domain. 
The main observation here is that correlations in velocity 
fluctuations stretch across the system for all packing 
fractions studied. We will come 
back to this point in a later section where we analyze systems 
of different sizes.
We also find that correlations are stronger for cubes than for a spheres at a given packing fraction. 

\subsubsection{Velocity auto-correlation}

\begin{figure}
    \centering
    \includegraphics[width=\linewidth]{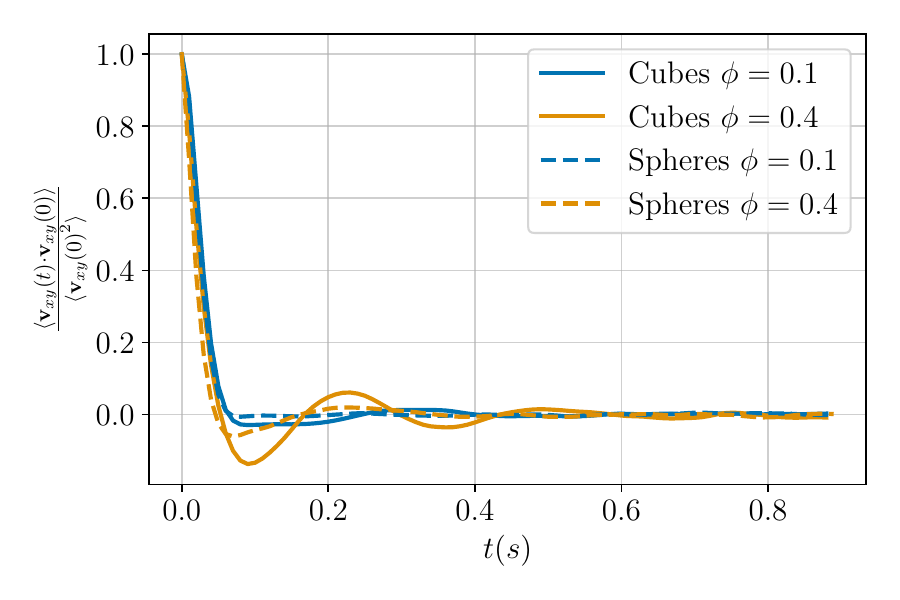}
    \caption{\textbf{Velocity-autocorrelation for cubes and spheres at two different packing fractions}.}
    \label{fig:vacf}
\end{figure}

We calculate velocity 
autocorrelation for hindered settling of cubes and spheres.  We calculate the velocity autocorrelation for $x$ and $y$ direction
\begin{equation}
C(t) = \frac{\langle \mathbf{v}_{\mathrm{xy}}(t) \cdot \mathbf{v}_{\mathrm{xy}}(0) \rangle}{\langle {\mathbf{v}_{\mathrm{xy}}(0)}^2 \rangle}.
\end{equation}
The results are given in Fig. \ref{fig:vacf}.
VACF of cubes at higher packing fraction shows a oscillating decay, similar to a under-damped oscillator. A similar observation is reported in \citet{fornariSedimentationFinitesizeSpheres2016} for the vertical component of velocity, which is a study carried out at much lower packing fractions and higher $Re$.
\subsection{System Size Dependence}
\begin{figure*}[t]
    \centering
    \includegraphics[width=\linewidth]{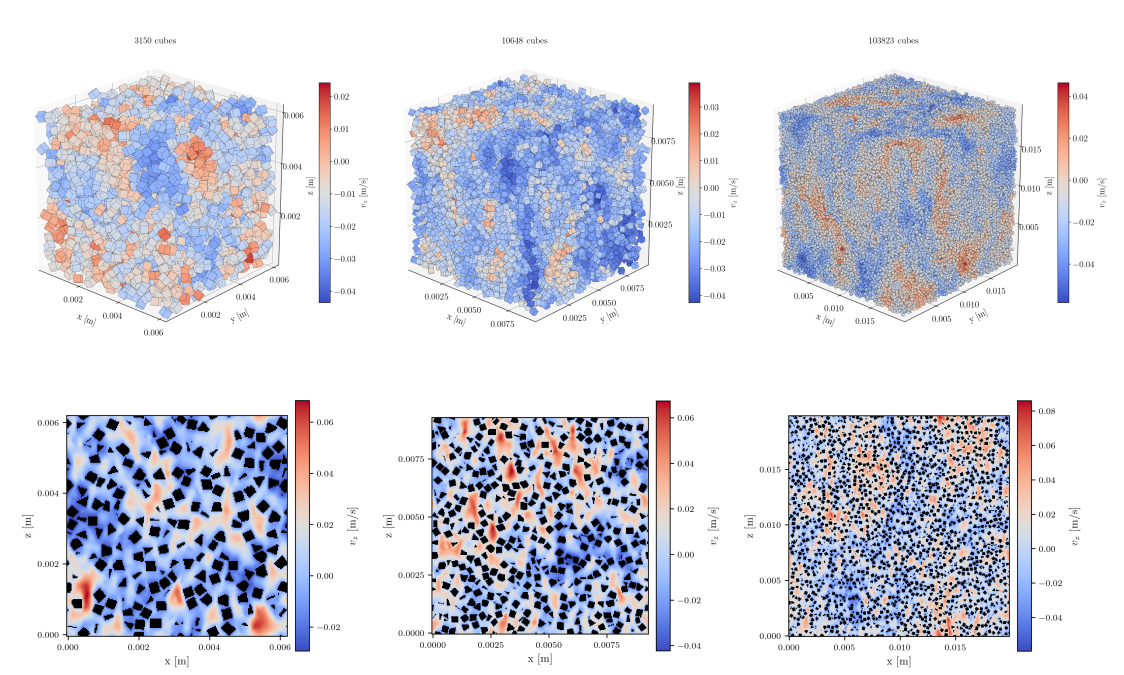}
    \caption{\textbf{Visualization of cubes undergoing hindered settling for three system sizes studied at $\phi=0.3$}. The top row shows the cubes colored by $v_z$ for $3150,10648$ and $103823$ cubes respectively. The bottom row are slices along the $xz$ plane showing fluid velocities along with the cubes for the corresponding system sizes.}
    \label{fig:visualization}
\end{figure*}
\begin{figure}[h]
    \centering
    \includegraphics[width=\linewidth]{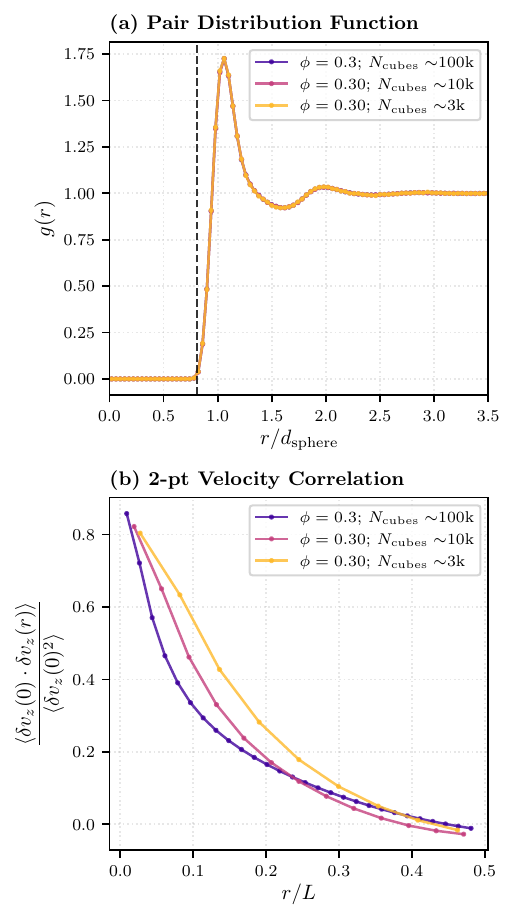}
    \caption{Correlation functions for different system sizes studied.} 
    \label{fig:systemSize}
\end{figure}
\begin{figure}[h]
    \centering
    \includegraphics[width=\linewidth]{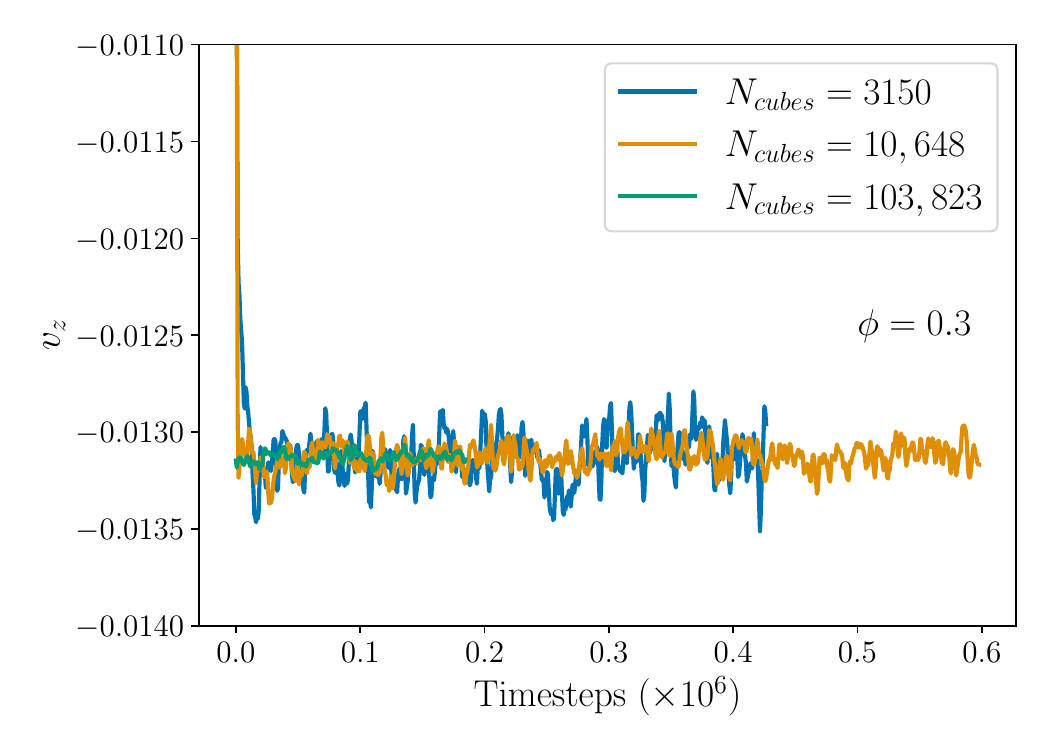}
    \caption{\textbf{Terminal velocity $v_z$ at $\phi=0.3$ for different system sizes}. The system size does not strongly affect the terminal velocity of bulk suspensions.}
    \label{fig:vzDiffSystems}
\end{figure}
\begin{table}[htbp]
  \centering
  \begin{tabular}{lrrr}
    \toprule
    \midrule
    \textbf{Particles}                  & \textbf{3{,}150}     & \textbf{10{,}648}  & \textbf{103{,}823} \\
    Lattice nodes (M)          & 9.53                 & 32.16                   & 314.43 \\
    Nodes per particle         & 3{,}025              & 3{,}020                 & 3{,}029 \\
    MPI ranks                  & 16                   & 64                      & 64 \\
    Approx Particles per rank         & 196.9                & 166.4                   & 1{,}622.2 \\
    Avg.\ MLUPs                & 24.30                & 111.58                  & 109.58 \\
    MLUPs per rank             & 1.52                 & 1.74                    & 1.71 \\
    \bottomrule
  \end{tabular}
    \caption{  \label{tab:cube_sed_scaling}
Parallel efficiency and scaling across three cube-sedimentation runs at different particle counts. All simulations are ran on AMD EPYC 7643 48-Core Processor.
} 
\end{table}
To demonstrate the capability of our framework, we study 
systems at $22^3=10,648$ and $47^3=103,823$ cube systems. We 
study both systems at a $\phi=0.3$. The correlation functions 
calculated from these simulations and the settling velocity are shown in Fig. \ref{fig:systemSize} and Fig \ref{fig:vzDiffSystems}. The primary conclusions 
from this study is that the average settling velocity and local 
arrangement of the cubes do not change as we study large 
systems. However, the 2-pt velocity correlation in space shows 
that the correlation length scales with the system size even for the largest system size we studied.

Previous works have studied dependence of velocity fluctuations on 
system size \citep{segreLongRangeCorrelationsSedimentation1997,guazzelliEvolutionParticlevelocityCorrelations2001,segreEffectiveGravitationalTemperature} 
experimentally and \citet{laddHydrodynamicInteractionsSuspension1988,laddSedimentationHomogeneousSuspensions1997} numerically. Spatial correlation in velocity 
fluctuations have been previously reported in 
\cite{guazzelliEvolutionParticlevelocityCorrelations2001,segreLongRangeCorrelationsSedimentation1997} experimentally and numerically in  
\citet{climentNumericalSimulationsRandom2003, uhlmannSedimentationDiluteSuspension2014,yaoEffectsParticleClustering2021}. As summarized in \citet{guazzelliFluctuationsInstabilitySedimentation2011} and \citet{guazzelliEvolutionParticlevelocityCorrelations2001}, the velocity fluctuations were 
found to saturate to a steady value as the box size exceeded $20 a \phi^{-1/3}$ in experiments. The initial fluctuations were found to be 
of the size of the container which would diminish with time and reach a 
size of $20a\phi^{-1/3}$. Numerically, the packing fraction $\phi$ and number of 
particles $N$ and the box dimensions are related as $L^3 = \frac{N\pi a^3}{6\phi}$. For $L^3> (2\times 20 a)^3 \phi^{-1}$, $N\gtrsim 
122,000$, which is the minimum size required to observe correlations at the discussed length-scale. 

Numerical observations from  
\citet{climentNumericalSimulationsRandom2003,uhlmannSedimentationDiluteSuspension2014,yaoEffectsParticleClustering2021} shows that the 
correlation in fluctuations along the vertical component of the velocity 
does not decay to zero even for the largest system, however the system 
sizes studied are not comparable to the limit observed in the 
experiments. The largest system size studied here approaches the limit 
and does not show a system that is uncorrelated spatially. 

As mentioned in \citet{laddEffectsContainerWalls2002,laddHydrodynamicScreeningSedimenting1996}, this could be an artifact of the periodic 
boundary conditions applied to top and bottom walls in the 
system. Details of different systems are given in Table \ref{tab:cube_sed_scaling}. A visualization of the three system sizes studied is shown in Fig. \ref{fig:visualization}
where we can see the system scale correlation in velocities.

\section{Conclusions}

We have implemented and validated a new framework to simulate 
irregular shaped particles in a fluid using open-source codes 
such as OpenLB and LAMMPS. We demonstrate the capabilities of 
our framework by simulating $\mathcal{O}(10^5)$ cubes in a 
fluid. This is an order of magnitude larger than \citet{studenikOpenHFDIBDEMExtensionOpenFOAM2024} where they 
demonstrated $\mathcal{O}(10^4)$ rigid bodies. We have used 
this capability to study hindered settling of spheres and 
cubes. Our analysis of local structure 
of cubes and spheres show qualitative agreement with 
previously published results in \citet{seyed-ahmadiSedimentationInertialMonodisperse2021,kunduSettlingDynamicsNonBrownian2025,fareedCollectiveSedimentationSymmetric2026,yinHinderedSettlingVelocity2007} although previous studies are at different $Re$'s. Spatial correlation in 
velocity fluctuations calculated at different system sizes 
indicate that the velocities remain correlated across the 
system even at the largest system studied. 
This problem is also addressed in \citet{moricheParticleresolvedSimulationsSettling2026}. 
This required further analysis and experimentation with different periodic boundary conditions, box dimensions and also perhaps the balancing force acting on the fluid nodes.

The framework can be further optimized in various ways. Currently, 
the implementation requires LAMMPS and OpenLB to have same 
number of MPI threads. An optimal number of threads for LAMMPS 
and OpenLB might be different at different densities which 
could make the code faster. GPU implementation of DEM 
simulations is also another direction for optimization. 
Future versions will include the lubrication force between 
particles. 

Future work will also be able to directly utilize the HLBM moving wall model \cite{kummerlaenderHFF2026} for turbulent flows of moving arbitrarily shaped particles.

\section {Acknowledgements}

DG and VB acknowledge support from the United States Army Corps of Engineers, Engineer Research and Development Center’s Cold Regions Research and Engineering Laboratory (ERDC-CRREL, Contract No. W913E524C0009). Any opinions, findings, and conclusions or recommendations expressed in this material are those of the author(s) and do not necessarily reflect the views of the Broad Agency Announcement Program and ERDC-CRREL. The authors would like to thank Stony Brook Research Computing and Cyberinfrastructure and the Institute for Advanced Computational Science at Stony Brook University for access to the high-performance SeaWulf computing system, which was made possible by \$1.85M in grants from the National Science Foundation (awards 1531492 and 2215987) and matching funds from the Empire State Development’s Division of Science, Technology, and Innovation (NYSTAR) program (contract C210148). We also thank Srikanth Sastry and Pingali Niharika Shankar for valuable discussions.

AK and MK acknowledge funding by the Deutsche Forschungsgemeinschaft (DFG, German Research Foundation) – project number 422374351.

\section{Declaration of generative AI and AI-assisted technologies in the manuscript preparation process.}
During the preparation of this work the author(s) used Google-Gemini and Claude-Code for formatting the manuscript and for generating python code involved in plotting the figures. After using this tool/service, the author(s) reviewed and edited the content as needed and take(s) full responsibility for the content of the published article.
\bibliographystyle{cas-model2-names}
\bibliography{CFD-DEM}

@article{guazzelliEvolutionParticlevelocityCorrelations2001,
  title = {Evolution of Particle-Velocity Correlations in Sedimentation},
  author = {Guazzelli, {\'E}lisabeth},
  year = {2001},
  journaltitle = {Physics of Fluids},
  shortjournal = {Phys. Fluids},
  doi = {10.1063/1.1369606},
  url = {https://sci-hub.ru/10.1063/1.1369606},
  urldate = {2026-07-24}
}

@article{hasimoto1959periodic,
  title={On the periodic fundamental solutions of the Stokes equations and their application to viscous flow past a cubic array of spheres},
  author={Hasimoto, Hidenori},
  journal={Journal of Fluid Mechanics},
  volume={5},
  number={2},
  pages={317--328},
  year={1959},
  publisher={Cambridge University Press}
}

@article{segreEffectiveGravitationalTemperature,
  title = {Effective {{Gravitational Temperature}} for {{Sedimentation}}},
  author = {Segr{\`e}, P N and Liu, F and Umbanhowar, P and Weitz, D A},
  langid = {english}
}

@book{krugerLatticeBoltzmannMethod2017a,
  title = {The {{Lattice Boltzmann Method}}: {{Principles}} and {{Practice}}},
  shorttitle = {The {{Lattice Boltzmann Method}}},
  author = {Kr{\"u}ger, Timm and Kusumaatmaja, Halim and Kuzmin, Alexandr and Shardt, Orest and Silva, Goncalo and Viggen, Erlend Magnus},
  year = 2017,
  series = {Graduate {{Texts}} in {{Physics}}},
  publisher = {Springer International Publishing},
  address = {Cham},
  doi = {10.1007/978-3-319-44649-3},
  urldate = {2026-07-28},
  copyright = {http://www.springer.com/tdm},
  isbn = {978-3-319-44647-9 978-3-319-44649-3},
  langid = {english}
}

@article{wen2014galilean,
  title={Galilean invariant fluid--solid interfacial dynamics in lattice Boltzmann simulations},
  author={Wen, Binghai and Zhang, Chaoying and Tu, Yusong and Wang, Chunlei and Fang, Haiping},
  journal={Journal of Computational Physics},
  volume={266},
  pages={161--170},
  year={2014},
  publisher={Elsevier}
}

@article{marquardtDiscreteContactModel2023,
  title = {A Discrete Contact Model for Complex Arbitrary-Shaped Convex Geometries},
  author = {Marquardt, Jan E. and Römer, Ulrich J. and Nirschl, Hermann and Krause, Mathias J.},
  date = {2023-09},
  journaltitle = {Particuology},
  volume = {80},
  pages = {180--191},
  issn = {16742001},
  doi = {10.1016/j.partic.2022.12.005},
  url = {https://linkinghub.elsevier.com/retrieve/pii/S1674200122002784},
  urldate = {2024-04-16},
  langid = {english}
}

@article{guazzelliFluctuationsInstabilitySedimentation2011,
  title = {Fluctuations and {{Instability}} in {{Sedimentation}}},
  author = {Guazzelli, {\'E}lisabeth and Hinch, John},
  year = 2011,
  month = jan,
  journal = {Annual Review of Fluid Mechanics},
  volume = {43},
  number = {1},
  pages = {97--116},
  issn = {0066-4189, 1545-4479},
  doi = {10.1146/annurev-fluid-122109-160736},
  urldate = {2026-06-15},
  langid = {english}
}

@article{yaoEffectsParticleClustering2021,
  title = {The Effects of Particle Clustering on Hindered Settling in High-Concentration Particle Suspensions},
  author = {Yao, Yinuo and Criddle, Craig S. and Fringer, Oliver B.},
  year = 2021,
  month = aug,
  journal = {Journal of Fluid Mechanics},
  volume = {920},
  pages = {A40},
  issn = {0022-1120, 1469-7645},
  doi = {10.1017/jfm.2021.470},
  urldate = {2026-07-22},
  langid = {english}
}

@article{laddSedimentationHomogeneousSuspensions1997,
  title = {Sedimentation of Homogeneous Suspensions of Non-{{Brownian}} Spheres},
  author = {Ladd, Anthony J. C.},
  year = 1997,
  month = mar,
  journal = {Physics of Fluids},
  volume = {9},
  number = {3},
  pages = {491--499},
  issn = {1070-6631, 1089-7666},
  doi = {10.1063/1.869212},
  urldate = {2026-06-15},
  langid = {english}
}

@article{moricheParticleresolvedSimulationsSettling2026,
  title = {Particle-Resolved Simulations of Settling Particles: A Methodology for Long Time-Integration Intervals},
  shorttitle = {Particle-Resolved Simulations of Settling Particles},
  author = {Moriche, M. and {Garc{\'i}a-Villalba}, M. and Uhlmann, M.},
  year = 2026,
  month = may,
  journal = {Acta Mechanica},
  issn = {0001-5970, 1619-6937},
  doi = {10.1007/s00707-026-04653-1},
  urldate = {2026-07-20},
  langid = {english}
}

@article{climentNumericalSimulationsRandom2003,
  title = {Numerical Simulations of Random Suspensions at Finite {{Reynolds}} Numbers},
  author = {Climent, E. and Maxey, M. R.},
  year = 2003,
  journal = {International Journal of Multiphase Flow},
  doi = {10.1016/s0301-9322(03)00016-8},
  urldate = {2026-07-15}
}

@article{fornariSedimentationFinitesizeSpheres2016,
  title = {Sedimentation of Finite-Size Spheres in Quiescent and Turbulent Environments},
  author = {Fornari, Walter and Picano, Francesco and Brandt, Luca},
  year = 2016,
  month = feb,
  journal = {Journal of Fluid Mechanics},
  volume = {788},
  pages = {640--669},
  issn = {0022-1120, 1469-7645},
  doi = {10.1017/jfm.2015.698},
  urldate = {2026-04-26},
  copyright = {https://www.cambridge.org/core/terms},
  langid = {english}
}

@article{laddEffectsContainerWalls2002,
  title = {Effects of {{Container Walls}} on the {{Velocity Fluctuations}} of {{Sedimenting Spheres}}},
  author = {Ladd, A. J. C.},
  year = 2002,
  month = jan,
  journal = {Physical Review Letters},
  volume = {88},
  number = {4},
  pages = {048301},
  issn = {0031-9007, 1079-7114},
  doi = {10.1103/PhysRevLett.88.048301},
  urldate = {2026-07-17},
  copyright = {http://link.aps.org/licenses/aps-default-license},
  langid = {english}
}

@article{garside1977velocity,
  title={Velocity-voidage relationships for fluidization and sedimentation in solid-liquid systems},
  author={Garside, John and Al-Dibouni, Maan R},
  journal={Industrial \& engineering chemistry process design and development},
  volume={16},
  number={2},
  pages={206--214},
  year={1977},
  publisher={ACS Publications}
}

@article{uhlmannSedimentationDiluteSuspension2014,
  title = {Sedimentation of a Dilute Suspension of Rigid Spheres at Intermediate {{Galileo}} Numbers: The Effect of Clustering upon the Particle Motion},
  shorttitle = {Sedimentation of a Dilute Suspension of Rigid Spheres at Intermediate {{Galileo}} Numbers},
  author = {Uhlmann, Markus and Doychev, Todor},
  year = 2014,
  month = aug,
  journal = {Journal of Fluid Mechanics},
  volume = {752},
  eprint = {1406.1667},
  primaryclass = {physics},
  pages = {310--348},
  issn = {0022-1120, 1469-7645},
  doi = {10.1017/jfm.2014.330},
  urldate = {2026-04-26},
  archiveprefix = {arXiv}
}

@article{laddHydrodynamicScreeningSedimenting1996,
  title = {Hydrodynamic {{Screening}} in {{Sedimenting Suspensions}} of Non-{{Brownian Spheres}}},
  author = {Ladd, Anthony J. C.},
  year = 1996,
  month = feb,
  journal = {Physical Review Letters},
  volume = {76},
  number = {8},
  pages = {1392--1395},
  issn = {0031-9007, 1079-7114},
  doi = {10.1103/PhysRevLett.76.1392},
  urldate = {2026-06-15},
  copyright = {http://link.aps.org/licenses/aps-default-license},
  langid = {english}
}

@article{batchelorSedimentationDiluteDispersion1972,
  title = {Sedimentation in a Dilute Dispersion of Spheres},
  author = {Batchelor, G. K.},
  year = 1972,
  month = mar,
  journal = {Journal of Fluid Mechanics},
  volume = {52},
  number = {2},
  pages = {245--268},
  issn = {0022-1120, 1469-7645},
  doi = {10.1017/S0022112072001399},
  urldate = {2026-07-15},
  copyright = {https://www.cambridge.org/core/terms},
  langid = {english}
}

@article{fareedCollectiveSedimentationSymmetric2026,
  title = {Collective Sedimentation of Symmetric Nonspherical Particles in {{Stokes}} Flow},
  author = {Fareed, Bilal and Nadeem, Muhammad and Ullah, Atta and Molina, John J. and Yamamoto, Ryoichi and Chamorro, Leonardo P. and Hamid, Adnan},
  year = 2026,
  month = may,
  journal = {Physical Review Fluids},
  volume = {11},
  number = {5},
  pages = {054305},
  issn = {2469-990X},
  doi = {10.1103/441b-p5mv},
  urldate = {2026-07-10},
  langid = {english}
}

@article{krauseParticleFlowSimulations2017,
	title = {Particle flow simulations with homogenised lattice {Boltzmann} methods},
	volume = {34},
	issn = {16742001},
	url = {https://linkinghub.elsevier.com/retrieve/pii/S167420011730041X},
	doi = {10.1016/j.partic.2016.11.001},
	language = {english},
	urldate = {2024-04-12},
	journal = {Particuology},
	author = {Krause, Mathias J. and Klemens, Fabian and Henn, Thomas and Trunk, Robin and Nirschl, Hermann},
	month = oct,
	year = {2017},
	pages = {1--13},
}

@phdthesis{najuchSimulationDenseSuspensions,
  title = {Simulation of Dense Suspensions with Discrete Element Method and a Coupled Lattice {{Boltzmann}} Method},
  author = {Najuch, Tim},
  langid = {english},
  year ={2019},
 publisher={The University of Edinburgh}
}

@article{thompsonLAMMPSFlexibleSimulation2022,
	title = {{LAMMPS} - a flexible simulation tool for particle-based materials modeling at the atomic, meso, and continuum scales},
	volume = {271},
	issn = {0010-4655},
	url = {https://www.sciencedirect.com/science/article/pii/S0010465521002836},
	doi = {10.1016/j.cpc.2021.108171},
	urldate = {2024-11-06},
	journal = {Computer Physics Communications},
	author = {Thompson, Aidan P. and Aktulga, H. Metin and Berger, Richard and Bolintineanu, Dan S. and Brown, W. Michael and Crozier, Paul S. and in 't Veld, Pieter J. and Kohlmeyer, Axel and Moore, Stan G. and Nguyen, Trung Dac and Shan, Ray and Stevens, Mark J. and Tranchida, Julien and Trott, Christian and Plimpton, Steven J.},
	month = feb,
	year = {2022},
	pages = {108171},
}

@article{olbPaper2021,
	title = {{OpenLB}{\textendash}{Open} source lattice {Boltzmann} code},
	volume = {81},
	issn = {0898-1221},
	url = {http://www.sciencedirect.com/science/article/pii/S0898122120301875},
	doi = {https://doi.org/10.1016/j.camwa.2020.04.033},
	journal = {Computers \& Mathematics with Applications},
	author = {Krause, M.J. and Kummerl{\"a}nder, A. and Avis, S.J. and Kusumaatmaja, H. and Dapelo, D. and Klemens, F. and Gaedtke, M. and Hafen, N. and Mink, A. and Trunk, R. and Marquardt, J.E. and Maier, M.L. and Haussmann, M. and Simonis, S.},
	year = {2021},
	pages = {258--288},
}

@article{marquardtReviewHomogenizedLattice2024,
	title = {A {Review} of the {Homogenized} {Lattice} {Boltzmann} {Method} for {Particulate} {Flow} {Simulations}: {From} {Fundamentals} to {Applications}},
	volume = {3},
	copyright = {http://creativecommons.org/licenses/by/3.0/},
	issn = {2674-0516},
	shorttitle = {A {Review} of the {Homogenized} {Lattice} {Boltzmann} {Method} for {Particulate} {Flow} {Simulations}},
	url = {https://www.mdpi.com/2674-0516/3/4/27},
	doi = {10.3390/powders3040027},
	language = {english},
	number = {4},
	urldate = {2025-07-30},
	journal = {Powders},
	publisher = {Multidisciplinary Digital Publishing Institute},
	author = {Marquardt, Jan E. and Krause, Mathias J.},
	month = dec,
	year = {2024},
	note = {Number: 4},
	pages = {500--530},
}

@article{marquardtNovelParticleDecomposition2024,
	title = {A novel particle decomposition scheme to improve parallel performance of fully resolved particulate flow simulations},
	volume = {78},
	issn = {1877-7503},
	url = {https://www.sciencedirect.com/science/article/pii/S1877750324000565},
	doi = {10.1016/j.jocs.2024.102263},
	urldate = {2025-08-22},
	journal = {Journal of Computational Science},
	author = {Marquardt, Jan E. and Hafen, Nicolas and Krause, Mathias J.},
	month = jun,
	year = {2024},
	pages = {102263},
}

@article{marquardtNovelModelDirect2024a,
	title = {A novel model for direct numerical simulation of suspension dynamics with arbitrarily shaped convex particles},
	volume = {304},
	issn = {0010-4655},
	url = {https://www.sciencedirect.com/science/article/pii/S0010465524002443},
	doi = {10.1016/j.cpc.2024.109321},
	urldate = {2025-08-22},
	journal = {Computer Physics Communications},
	author = {Marquardt, Jan E. and Hafen, Nicolas and Krause, Mathias J.},
	month = nov,
	year = {2024},
	pages = {109321},
}

@article{ahmadianSimulatingFluidSolid2024,
	title = {Simulating the fluid{\textendash}solid interaction of irregularly shaped particles using the {LBM}-{DEM} coupling method},
	volume = {171},
	issn = {0266352X},
	url = {https://linkinghub.elsevier.com/retrieve/pii/S0266352X24003318},
	doi = {10.1016/j.compgeo.2024.106395},
	language = {english},
	urldate = {2025-09-01},
	journal = {Computers and Geotechnics},
	author = {Ahmadian, Mohammad Hassan and Zheng, Wenbo},
	month = jul,
	year = {2024},
	pages = {106395},
}

@article{trunkRevisitingHomogenizedLattice2021,
	title = {Revisiting the {Homogenized} {Lattice} {Boltzmann} {Method} with {Applications} on {Particulate} {Flows}},
	volume = {9},
	copyright = {http://creativecommons.org/licenses/by/3.0/},
	issn = {2079-3197},
	url = {https://www.mdpi.com/2079-3197/9/2/11},
	doi = {10.3390/computation9020011},
	language = {english},
	number = {2},
	urldate = {2025-09-05},
	journal = {Computation},
	publisher = {Multidisciplinary Digital Publishing Institute},
	author = {Trunk, Robin and Weckerle, Timo and Hafen, Nicolas and Th{\"a}ter, Gudrun and Nirschl, Hermann and Krause, Mathias J.},
	month = feb,
	year = {2021},
	pages = {11},
}

@article{tencateParticleImagingVelocimetry2002,
	title = {Particle imaging velocimetry experiments and lattice-{Boltzmann} simulations on a single sphere settling under gravity},
	volume = {14},
	issn = {1070-6631, 1089-7666},
	url = {https://pubs.aip.org/pof/article/14/11/4012/1008754/Particle-imaging-velocimetry-experiments-and},
	doi = {10.1063/1.1512918},
	language = {english},
	number = {11},
	urldate = {2025-09-13},
	journal = {Physics of Fluids},
	author = {Ten Cate, A. and Nieuwstad, C. H. and Derksen, J. J. and Van Den Akker, H. E. A.},
	month = nov,
	year = {2002},
	pages = {4012--4025},
}

@article{liContactModelNormal2012,
	title = {A contact model for normal immersed collisions between a particle and a wall},
	volume = {691},
	copyright = {https://www.cambridge.org/core/terms},
	issn = {0022-1120, 1469-7645},
	url = {https://www.cambridge.org/core/product/identifier/S0022112011004617/type/journal_article},
	doi = {10.1017/jfm.2011.461},
	language = {english},
	urldate = {2025-09-13},
	journal = {Journal of Fluid Mechanics},
	author = {Li, Xiaobai and Hunt, Melany L. and Colonius, Tim},
	month = jan,
	year = {2012},
	pages = {123--145},
}

@article{laddNumericalSimulationsParticulate1994,
	title = {Numerical simulations of particulate suspensions via a discretized {Boltzmann} equation. {Part} 1. {Theoretical} foundation},
	volume = {271},
	copyright = {https://www.cambridge.org/core/terms},
	issn = {0022-1120, 1469-7645},
	url = {https://www.cambridge.org/core/product/identifier/S0022112094001771/type/journal_article},
	doi = {10.1017/S0022112094001771},
	language = {english},
	urldate = {2026-01-21},
	journal = {Journal of Fluid Mechanics},
	author = {Ladd, Anthony J. C.},
	month = jul,
	year = {1994},
	pages = {285--309},
}

@article{laddNumericalSimulationsParticulate1994a,
	title = {Numerical simulations of particulate suspensions via a discretized {Boltzmann} equation. {Part} 2. {Numerical} results},
	volume = {271},
	copyright = {https://www.cambridge.org/core/terms},
	issn = {0022-1120, 1469-7645},
	url = {https://www.cambridge.org/core/product/identifier/S0022112094001783/type/journal_article},
	doi = {10.1017/S0022112094001783},
	language = {english},
	urldate = {2026-02-02},
	journal = {Journal of Fluid Mechanics},
	author = {Ladd, Anthony J. C.},
	month = jul,
	year = {1994},
	pages = {311--339},
}

@article{laddHydrodynamicInteractionsSuspension1988,
	title = {Hydrodynamic interactions in a suspension of spherical particles},
	url = {https://sci-hub.ru/10.1063/1.454658},
	doi = {10.1063/1.454658},
	urldate = {2026-02-27},
	journal = {The Journal of Chemical Physics},
	author = {Ladd},
	year = {1988},
}

@article{nobleLatticeBoltzmannMethodPartially1998,
  title = {A Lattice-{{Boltzmann}} Method for Partially Saturated Computational Cells},
  author = {Noble, {\relax DR} and Torczynski, {\relax JR}},
  year = 1998,
  journal = {International Journal of Modern Physics C},
  volume = {9},
  number = {08},
  pages = {1189--1201},
  publisher = {World Scientific}
}

@article{rettingerComparativeStudyFluidparticle2017,
	title = {A comparative study of fluid-particle coupling methods for fully resolved lattice {Boltzmann} simulations},
	volume = {154},
	issn = {00457930},
	url = {http://arxiv.org/abs/1702.04910},
	doi = {10.1016/j.compfluid.2017.05.033},
	urldate = {2026-03-03},
	journal = {Computers \& Fluids},
	author = {Rettinger, Christoph and R{\"u}de, Ulrich},
	month = sep,
	year = {2017},
	note = {arXiv:1702.04910 [cs]},
	pages = {74--89},
}

@misc{rettingerCoupledLatticeBoltzmann2017,
	title = {A {Coupled} {Lattice} {Boltzmann} {Method} and {Discrete} {Element} {Method} for {Discrete} {Particle} {Simulations} of {Particulate} {Flows}},
	url = {http://arxiv.org/abs/1711.00336},
	doi = {10.48550/arXiv.1711.00336},
	urldate = {2026-03-05},
	publisher = {arXiv},
	author = {Rettinger, Christoph and R{\"u}de, Ulrich},
	month = nov,
	year = {2017},
	note = {arXiv:1711.00336 [cs]},
}

@article{zhangLatticeBoltzmannSimulations2016,
	title = {Lattice {Boltzmann} simulations of settling behaviors of irregularly shaped particles},
	volume = {93},
	copyright = {http://link.aps.org/licenses/aps-default-license},
	issn = {2470-0045, 2470-0053},
	url = {https://link.aps.org/doi/10.1103/PhysRevE.93.062612},
	doi = {10.1103/PhysRevE.93.062612},
	language = {english},
	number = {6},
	urldate = {2026-03-22},
	journal = {Physical Review E},
	author = {Zhang, Pei and Galindo-Torres, S. A. and Tang, Hongwu and Jin, Guangqiu and Scheuermann, A. and Li, Ling},
	month = jun,
	year = {2016},
	pages = {062612},
}

@article{goniva2012influence,
  title={Influence of rolling friction on single spout fluidized bed simulation},
  author={Goniva, Christoph and Kloss, Christoph and Deen, Niels G and Kuipers, Johannes AM and Pirker, Stefan},
  journal={Particuology},
  volume={10},
  number={5},
  pages={582--591},
  year={2012},
  publisher={Elsevier}
  }

@techreport{syamlal1993mfix,
  title={MFIX documentation theory guide},
  author={Syamlal, Madhava and Rogers, William and OBrien, Thomas J},
  year={1993},
  institution={USDOE Morgantown Energy Technology Center, WV (United States)}
}

@article{blaisLetheOpensourceParallel2020,
  title = {Lethe: {{An}} Open-Source Parallel High-Order Adaptative {{CFD}} Solver for Incompressible Flows},
  shorttitle = {Lethe},
  author = {Blais, Bruno and Barbeau, Lucka and Bibeau, Val{\'e}rie and Gauvin, Simon and Geitani, Toni El and Golshan, Shahab and Kamble, Rajeshwari and Mirakhori, Ghazaleh and Chaouki, Jamal},
  year = 2020,
  month = jul,
  journal = {SoftwareX},
  volume = {12},
  pages = {100579},
  issn = {23527110},
  doi = {10.1016/j.softx.2020.100579},
  urldate = {2026-05-09},
  langid = {english}
}

@article{fluent2023ansys,
  title={Ansys fluent theory guide},
  author={Fluent, ANSYS},
  journal={(No Title)},
  year={2023}
}

@article{lattPalabosParallelLattice2021,
  title = {Palabos: {{Parallel Lattice Boltzmann Solver}}},
  shorttitle = {Palabos},
  author = {Latt, Jonas and Malaspinas, Orestis and Kontaxakis, Dimitrios and Parmigiani, Andrea and Lagrava, Daniel and Brogi, Federico and Belgacem, Mohamed Ben and Thorimbert, Yann and Leclaire, S{\'e}bastien and Li, Sha and Marson, Francesco and Lemus, Jonathan and Kotsalos, Christos and Conradin, Rapha{\"e}l and Coreixas, Christophe and Petkantchin, R{\'e}my and Raynaud, Franck and Beny, Jo{\"e}l and Chopard, Bastien},
  year = 2021,
  month = jan,
  journal = {Computers \& Mathematics with Applications},
  series = {Development and {{Application}} of {{Open-source Software}} for {{Problems}} with {{Numerical PDEs}}},
  volume = {81},
  pages = {334--350},
  issn = {0898-1221},
  doi = {10.1016/j.camwa.2020.03.022},
  urldate = {2026-05-09}
}

@article{najuchAnalysisTwoPartiallysaturatedcell2019a,
  title = {Analysis of Two Partially-Saturated-Cell Methods for Lattice {{Boltzmann}} Simulation of Granular Suspension Rheology},
  author = {Najuch, Tim and Sun, Jin},
  year = 2019,
  month = jul,
  journal = {Computers \& Fluids},
  volume = {189},
  pages = {1--12},
  issn = {00457930},
  doi = {10.1016/j.compfluid.2019.05.004},
  urldate = {2026-05-09},
  langid = {english}
}

@misc{maggio-aprileLEDDSPortableLBMDEM2025,
  title = {{{LEDDS}}: {{Portable LBM-DEM}} Simulations on {{GPUs}}},
  shorttitle = {{{LEDDS}}},
  author = {{Maggio-Aprile}, Raphael and Rambosson, Maxime and Coreixas, Christophe and Latt, Jonas},
  year = 2025,
  publisher = {arXiv},
  doi = {10.48550/ARXIV.2512.04997},
  urldate = {2026-05-09},
  copyright = {Creative Commons Attribution 4.0 International}
}

@article{capozzaHierarchicalSphericalHarmonicbased2021,
  title = {A Hierarchical, Spherical Harmonic-Based Approach to Simulate Abradable, Irregularly Shaped Particles in {{DEM}}},
  author = {Capozza, R. and Hanley, K.J.},
  year = 2021,
  month = jan,
  journal = {Powder Technology},
  volume = {378},
  pages = {528--537},
  issn = {00325910},
  doi = {10.1016/j.powtec.2020.10.015},
  urldate = {2026-05-09},
  langid = {english}
}

@article{luDiscreteElementModels2015,
  title = {Discrete Element Models for Non-Spherical Particle Systems: {{From}} Theoretical Developments to Applications},
  shorttitle = {Discrete Element Models for Non-Spherical Particle Systems},
  author = {Lu, G. and Third, J. R. and M{\"u}ller, C. R.},
  year = 2015,
  journal = {Chemical Engineering Science},
  doi = {10.1016/j.ces.2014.11.050},
  urldate = {2026-05-09}
}

@article{bauerWaLBerlaBlockstructuredHighperformance2021,
  title = {{{waLBerla}}: {{A}} Block-Structured High-Performance Framework for Multiphysics Simulations},
  shorttitle = {{{waLBerla}}},
  author = {Bauer, Martin and Eibl, Sebastian and Godenschwager, Christian and Kohl, Nils and Kuron, Michael and Rettinger, Christoph and Schornbaum, Florian and Schwarzmeier, Christoph and Th{\"o}nnes, Dominik and K{\"o}stler, Harald and R{\"u}de, Ulrich},
  year = 2021,
  month = jan,
  journal = {Computers \& Mathematics with Applications},
  series = {Development and {{Application}} of {{Open-source Software}} for {{Problems}} with {{Numerical PDEs}}},
  volume = {81},
  pages = {478--501},
  issn = {0898-1221},
  doi = {10.1016/j.camwa.2020.01.007},
  urldate = {2026-05-10}
}

@article{willenContinuityWavesResolvedparticle2017,
  title = {Continuity Waves in Resolved-Particle Simulations of Fluidized Beds},
  author = {Willen, Daniel P. and Sierakowski, Adam J. and Zhou, Gedi and Prosperetti, Andrea},
  year = 2017,
  month = nov,
  journal = {Physical Review Fluids},
  volume = {2},
  number = {11},
  pages = {114305},
  issn = {2469-990X},
  doi = {10.1103/PhysRevFluids.2.114305},
  urldate = {2026-05-10},
  copyright = {https://link.aps.org/licenses/aps-default-license},
  langid = {english}
}

@article{chongEffectParticleShape1979,
  title = {Effect of Particle Shape on Hindered Settling in Creeping Flow},
  author = {Chong, Y. S. and Ratkowsky, D. A. and Epstein, N.},
  year = 1979,
  journal = {Powder Technology},
  doi = {10.1016/0032-5910(79)85025-1},
  urldate = {2026-05-10}
}

@article{zhaoRevolutionizingGranularMatter2023,
  title = {Revolutionizing Granular Matter Simulations by High-Performance Ray Tracing Discrete Element Method for Arbitrarily-Shaped Particles},
  author = {Zhao, Shiwei and Zhao, Jidong},
  year = 2023,
  month = nov,
  journal = {Computer Methods in Applied Mechanics and Engineering},
  volume = {416},
  pages = {116370},
  issn = {00457825},
  doi = {10.1016/j.cma.2023.116370},
  urldate = {2026-05-09},
  langid = {english}
}

@inproceedings{seil2016lbdemcoupling,
  title={Lbdemcoupling: Open-source power for fluid-particle systems},
  author={Seil, Philippe and Pirker, Stefan},
  booktitle={International conference on discrete element methods},
  pages={679--686},
  year={2016},
  organization={Springer}
}

@article{studenikOpenHFDIBDEMExtensionOpenFOAM2024,
  title = {{{OpenHFDIB-DEM}}: {{An}} Extension to {{OpenFOAM}} for {{CFD-DEM}} Simulations with Arbitrary Particle Shapes},
  shorttitle = {{{OpenHFDIB-DEM}}},
  author = {Studen{\'i}k, Ond{\v r}ej and Isoz, Martin and Kotou{\v c} {\v S}ourek, Martin and Ko{\v c}{\'i}, Petr},
  year = 2024,
  month = sep,
  journal = {SoftwareX},
  volume = {27},
  pages = {101871},
  issn = {23527110},
  doi = {10.1016/j.softx.2024.101871},
  urldate = {2026-05-08},
  langid = {english}
}

@article{weller1998tensorial,
  title={A tensorial approach to computational continuum mechanics using object-oriented techniques},
  author={Weller, Henry G and Tabor, Gavin and Jasak, Hrvoje and Fureby, Christer},
  journal={Computers in physics},
  volume={12},
  number={6},
  pages={620--631},
  year={1998},
  publisher={American Institute of Physics}
}

@article{kloss2012models,
  title={Models, algorithms and validation for opensource DEM and CFD--DEM},
  author={Kloss, Christoph and Goniva, Christoph and Hager, Alice and Amberger, Stefan and Pirker, Stefan},
  journal={Progress in Computational Fluid Dynamics, an International Journal},
  volume={12},
  number={2-3},
  pages={140--152},
  year={2012},
  publisher={Inderscience Publishers}
}

@article{smilauer2023yade,
  title={Yade documentation},
  author={Smilauer, Vaclav and Angelidakis, Vasileios and Catalano, Emanuele and Caulk, Robert and Chareyre, Bruno and Chevremont, William and Dorofeenko, Sergei and Duriez, Jerome and Dyck, Nolan and Elias, Jan and others},
  journal={arXiv preprint arXiv:2301.00611},
  year={2023}
}

@article{kwonExperimentalInvestigationSettling2025,
	title = {Experimental investigation of settling velocity for cuboidal microplastic},
	volume = {161},
	issn = {0141-1187},
	url = {https://www.sciencedirect.com/science/article/pii/S0141118725002792},
	doi = {10.1016/j.apor.2025.104693},
	urldate = {2026-04-11},
	journal = {Applied Ocean Research},
	author = {Kwon, Seung-Jae and Park, Young-Gyu and Quyen, Le Duc and Lee, In-Cheol and Choi, Jun Myoung},
	month = aug,
	year = {2025},
	pages = {104693},
}

@article{yinHinderedSettlingVelocity2007,
	title = {Hindered settling velocity and microstructure in suspensions of solid spheres with moderate {Reynolds} numbers},
	url = {https://sci-hub.ru/10.1063/1.2764109},
	doi = {10.1063/1.2764109},
	urldate = {2026-04-14},
	journal = {Physics of Fluids},
	author = {Yin, Xiaolong and Koch, Donald L.},
	year = {2007},
}

@article{haiderDragCoefficientTerminal1989,
	title = {Drag coefficient and terminal velocity of spherical and nonspherical particles},
	url = {https://sci-hub.ru/10.1016/0032-5910(89)80008-7},
	doi = {10.1016/0032-5910(89)80008-7},
	urldate = {2026-04-17},
	journal = {Powder Technology},
	author = {Haider, A. and Levenspiel, O.},
	year = {1989},
}

@article{sanganiSlowFlowPeriodic1982,
	title = {Slow flow through a periodic array of spheres},
	url = {https://sci-hub.ru/10.1016/0301-9322(82)90047-7},
	doi = {10.1016/0301-9322(82)90047-7},
	urldate = {2026-04-18},
	journal = {International Journal of Multiphase Flow},
	author = {Sangani, A. S. and Acrivos, A.},
	year = {1982},
}

@article{richardsonSedimentationSuspensionUniform1954,
	title = {The sedimentation of a suspension of uniform spheres under conditions of viscous flow},
	volume = {3},
	issn = {0009-2509},
	url = {https://www.sciencedirect.com/science/article/pii/0009250954850159},
	doi = {10.1016/0009-2509(54)85015-9},
	number = {2},
	urldate = {2026-04-19},
	journal = {Chemical Engineering Science},
	author = {Richardson, J. F. and Zaki, W. N.},
	month = apr,
	year = {1954},
	pages = {65--73},
}

@article{paulInfluenceShapeSurface2017,
	title = {Influence of shape and surface charge on the sedimentation of spheroidal, cubic and rectangular cuboid particles},
	volume = {322},
	issn = {00325910},
	url = {https://linkinghub.elsevier.com/retrieve/pii/S0032591017307234},
	doi = {10.1016/j.powtec.2017.09.002},
	language = {english},
	urldate = {2026-04-19},
	journal = {Powder Technology},
	author = {Paul, Neepa and Biggs, Simon and Shiels, Jessica and Hammond, Robert B. and Edmondson, Michael and Maxwell, Lisa and Harbottle, David and Hunter, Timothy N.},
	month = dec,
	year = {2017},
	pages = {75--83},
}

@misc{kunduSettlingDynamicsNonBrownian2025,
	title = {Settling dynamics of non-{Brownian} suspension of spherical and cubic particles in {Stokes} flow},
	url = {http://arxiv.org/abs/2501.08091},
	doi = {10.48550/arXiv.2501.08091},
	urldate = {2026-04-19},
	publisher = {arXiv},
	author = {Kundu, Dipankar and Usabiaga, Florencio Balboa and Ellero, Marco},
	month = mar,
	year = {2025},
	note = {arXiv:2501.08091 [physics]},
}

@article{angelidakisCLUMPCodeLibrary2021,
	title = {{CLUMP}: {A} {Code} {Library} to generate {Universal} {Multi}-sphere {Particles}},
	volume = {15},
	issn = {2352-7110},
	shorttitle = {{CLUMP}},
	url = {https://www.sciencedirect.com/science/article/pii/S2352711021000704},
	doi = {10.1016/j.softx.2021.100735},
	urldate = {2026-04-28},
	journal = {SoftwareX},
	author = {Angelidakis, Vasileios and Nadimi, Sadegh and Otsubo, Masahide and Utili, Stefano},
	month = jul,
	year = {2021},
	pages = {100735},
}

@article{seyed-ahmadiSedimentationInertialMonodisperse2021,
	title = {Sedimentation of inertial monodisperse suspensions of cubes and spheres},
	volume = {6},
	issn = {2469-990X},
	url = {https://link.aps.org/doi/10.1103/PhysRevFluids.6.044306},
	doi = {10.1103/PhysRevFluids.6.044306},
	language = {english},
	number = {4},
	urldate = {2026-04-28},
	journal = {Physical Review Fluids},
	author = {Seyed-Ahmadi, Arman and Wachs, Anthony},
	month = apr,
	year = {2021},
	pages = {044306},
}

@article{segreLongRangeCorrelationsSedimentation1997,
	title = {Long-{Range} {Correlations} in {Sedimentation}},
	volume = {79},
	copyright = {http://link.aps.org/licenses/aps-default-license},
	issn = {0031-9007, 1079-7114},
	url = {https://link.aps.org/doi/10.1103/PhysRevLett.79.2574},
	doi = {10.1103/PhysRevLett.79.2574},
	language = {english},
	number = {13},
	urldate = {2026-04-28},
	journal = {Physical Review Letters},
	author = {Segr{\`e}, P. N. and Herbolzheimer, E. and Chaikin, P. M.},
	month = sep,
	year = {1997},
	pages = {2574--2577},
}

@article{cundallDiscreteNumericalModel1979,
	title = {A discrete numerical model for granular assemblies},
	volume = {29},
	issn = {0016-8505},
	url = {https://doi.org/10.1680/geot.1979.29.1.47},
	doi = {10.1680/geot.1979.29.1.47},
	number = {1},
	urldate = {2026-05-08},
	journal = {G{\'e}otechnique},
	author = {Cundall, P. A. and Strack, O. D. L.},
	month = mar,
	year = {1979},
	pages = {47--65},
}

@article{kummerlaenderHFF2026,
    author = {Kummerländer, Adrian and Ito, Shota and Schecher, Maximilian and Dapelo, Davide and Simonis, Stephan and Krause, Mathias J. and Bukreev, Fedor},
    title = {Efficient wall-modelled large eddy simulation of rotors using homogenized lattice Boltzmann methods},
    journal = {International Journal of Numerical Methods for Heat \& Fluid Flow},
    pages = {1-25},
    year = {2026},
    month = {04},
    issn = {0961-5539},
    doi = {10.1108/HFF-09-2025-0724},
    url = {https://doi.org/10.1108/HFF-09-2025-0724}}

@article{kummerlaenderCMAME2026,
title = {Efficient fluid structure interaction simulation of vocal fold oscillations using a homogenized Lattice Boltzmann Method},
journal = {Computer Methods in Applied Mechanics and Engineering},
volume = {457},
pages = {119009},
year = {2026},
issn = {0045-7825},
doi = {https://doi.org/10.1016/j.cma.2026.119009},
url = {https://www.sciencedirect.com/science/article/pii/S0045782526002823},
author = {Adrian Kummerländer and Bogac Tur and Maik Haase and Fedor Bukreev and Michael Döllinger and Mathias J. Krause and Stefan Kniesburges}
}

@article{angotPenalizationMethod1999,
  title = {A Penalization Method to Take into Account Obstacles in Incompressible Viscous Flows},
  author = {Angot, Philippe and Bruneau, Charles-Henri and Fabrie, Pierre},
  year = 1999,
  month = feb,
  journal = {Numerische Mathematik},
  shortjournal = {Numer. Math.},
  volume = {81},
  number = {4},
  pages = {497--520},
  issn = {0945-3245},
  doi = {10.1007/s002110050401},
  langid = {english}
}

\clearpage
\newpage
\appendix
\newpage
\setcounter{figure}{0}
\renewcommand{\thefigure}{A.\arabic{figure}}

\section{Methods details}
\label{MethodsDetails}
\subsection{Lattice Boltzmann Method}




\subsubsection{Target equations and volume penalization}

The fluid is governed by the incompressible Navier-Stokes equations on the
time-dependent fluid domain $\Omega_{\mathrm{f}}(t)=\Omega\setminus\bigcup_{\mathrm{k}}B_{\mathrm{k}}(t)$,
where $B_{\mathrm{k}}(t)$ is the region occupied by the rigid body $k$ at time $t$,
\begin{align}
\nabla\cdot\mathbf{u} & = 0, \label{eq:nseMass}\\
\frac{\partial\mathbf{u}}{\partial t} + \mathbf{u}\cdot\nabla\mathbf{u}
& = -\frac{\nabla p}{\rho} + \nu\nabla^{2}\mathbf{u}, \label{eq:nseMomentum}
\end{align}
subject to the no-slip condition
\begin{equation}
\mathbf{u} = \mathbf{u}_{\mathrm{p}} \quad \text{on } \partial B_{\mathrm{k}}(t),
\label{eq:noSlip}
\end{equation}
on every moving body surface, where $\mathbf{u}_{\mathrm{p}}$ is the local
velocity of the body. Rather than discretizing Eqs.
\ref{eq:nseMass}-\ref{eq:noSlip} directly, we solve a
penalized problem on the full, fixed domain $\Omega$,
\begin{equation}
\frac{\partial\mathbf{u}}{\partial t} + \mathbf{u}\cdot\nabla\mathbf{u}
= -\frac{\nabla p}{\rho} + \nu\nabla^{2}\mathbf{u}
  - \frac{1}{\eta_{\mathrm{p}}}\chi(\mathbf{x},t)\left(\mathbf{u}-\mathbf{u}_{\mathrm{p}}\right),
\label{eq:penalized}
\end{equation}
where $\chi(\mathbf{x},t)$ is the indicator function of the region occupied by
the bodies and $\eta_{\mathrm{p}}>0$ is the penalization parameter. The added
term is a Brinkman type volume penalization
\citep{angotPenalizationMethod1999}, formally the Darcy
drag of a porous medium of permeability $K=\nu\eta_{\mathrm{p}}$, although no
physical porous medium is modelled here. It vanishes in the fluid and drives the
velocity towards $\mathbf{u}_{\mathrm{p}}$ inside the bodies, so that solutions
of Eq. \ref{eq:penalized} converge to the solution of Eqs.
\ref{eq:nseMass}-\ref{eq:noSlip} for $\eta_{\mathrm{p}}\to0$
\citep{angotPenalizationMethod1999}. The geometry of the bodies, their motion and
their contacts enter only through $\chi$. The same formulation is used in
\citep{kummerlaenderCMAME2026} for the closing contact of oscillating vocal folds
and in \citep{kummerlaenderHFF2026} for blade-resolved simulations of wind turbine
rotors. The discrete
counterpart of $1-\chi$ is the porosity field $d(\mathbf{x},t)$ used throughout this
work, with $d=1$ on fluid nodes, $d=0$ on solid nodes and a continuous transition
between the two across the surface of a body.

\subsubsection{Lattice Boltzmann discretization}

The Lattice Boltzmann method solves Eqs. \ref{eq:nseMass}-\ref{eq:nseMomentum} via a
discrete in space-time kinetic theory. A discrete velocity set
${\cal C}=\{{\bf C}_1,\cdots, {\bf C}_n\}$ with associated populations
$\{f_1,\cdots, f_n\}$ and non-negative weights $\{w_1,\cdots, w_n\}$ is chosen to
respect the symmetries required by the fluid flow. Grid spacing $\Delta x$ and time
step $\Delta t$ are chosen such that
particle populations move precisely from one
lattice node to an adjacent node along the discrete
velocity vectors $\mathbf{C}_i$ in a single time step
$\Delta t$. The $f_i$ populations
are related to density $\rho$ and velocity $\mathbf{v}$ of the fluid
at a lattice node as

\begin{align*}
\rho(\mathbf{x},t) & =\sum_{\mathrm{i}}^{n}f_{\mathrm{i}}(\mathbf{x},t)\\
\mathbf{\rho v}(\mathbf{x},t & )=\sum_{\mathrm{i}}^{n}\mathbf{C}_{\mathrm{i}}f_{\mathrm{i}}(\mathbf{x},t)
\end{align*}
. The variable $f_{\mathrm{i}}$'s are evolved using the following equation
\begin{align}
f_{\mathrm{i}}(\mathbf{x}+\mathbf{C}_{\mathrm{i}}\Delta t,t+\Delta t) & =f_{\mathrm{i}}(\mathbf{x},t)+\frac{1}{\tau}\left[f_{\mathrm{i}}^{eq}(\rho,\mathbf{v})-f_{\mathrm{i}}(\mathbf{x}.t)\right]\label{eq:LBMupdate}
\end{align}
where 
\[
f_{\mathrm{i}}^{eq}(\rho,\mathbf{v})=w_{\mathrm{i}}\rho\times\left(1+\frac{\mathbf{C}_{\mathrm{i}}.\mathbf{v}}{c_{\mathrm{s}}^{2}}+\frac{(\mathbf{C}_{\mathrm{i}}.\mathbf{v})^{2}}{2c_{\mathrm{s}}^{4}}-\frac{\mathbf{v}^{2}}{2c_{\mathrm{s}}^{2}}\right)
\]
is the equilibrium distribution. Here $c_{\mathrm{s}}$ is a constant with
dimension of velocity. The simulation consists of two steps : a collision
step where RHS of eq. \ref{eq:LBMupdate} is calculated at a lattice
node and a streaming step where the updated $f_{\mathrm{i}}$'s are moved or
streamed to the adjacent lattice node connected by the vector $\mathbf{C}_{\mathrm{i}}$.
A Chapman-Enskog analysis, as shown in various standard textbooks
\citep{krugerLatticeBoltzmannMethod2017a}, recovers
Eqs. \ref{eq:nseMass}-\ref{eq:nseMomentum} with the kinematic viscosity
\begin{equation}
\nu=c_{\mathrm{s}}^{2}\left(\tau-\frac{1}{2}\right) \frac{\Delta x^2}{\Delta t}\label{eq:viscosity}
\end{equation}
so that $\tau>0.5$ is required for physical viscosities. We use the D3Q19 velocity
set, for which mass and momentum are conserved. The penalization term of
Eq. \ref{eq:penalized} is recovered by adding a forcing term to the collision step,
as described in Appendix \ref{FSIdetails}.

\subsection{DEM}

First we will discuss the DEM method used in the simulations for particles. We will discuss the contact force between spherical particles
of different radii. Since we are using the clump
representation for arbitrary shapes, all interactions between rigid
bodies can be resolved through interaction between spheres. The force
acting between two spheres $i$ and $j$ of radii $R_{\mathrm{i}}$ and $R_{\mathrm{j}}$
is given by a normal component, 

\[
\mathbf{F}_{\mathrm{n}}=\frac{4}{3}E_{\mathrm{eff}}R^{1/2}\delta_{\mathrm{ij}}^{3/2}\mathbf{n}-\eta_{\mathrm{n}}\mathbf{\boldsymbol{v}}_{\mathrm{n,rel}}
\]
which is a repulsive force acting when particles overlap. Here $E_{\mathrm{eff}}$
is the effective Young's modulus given by 
\[
E_{\mathrm{eff}}=\frac{1}{\left(\frac{1-\nu_{\mathrm{i}}^{2}}{E_{\mathrm{i}}}+\frac{1-\nu_{\mathrm{j}}^{2}}{E_{\mathrm{j}}}\right)}
\]
$,R=\frac{R_{\mathrm{i}}R_{\mathrm{j}}}{R_{\mathrm{i}}+R_{\mathrm{j}}}$, $\delta_{\mathrm{ij}}=R_{\mathrm{i}}+R_{\mathrm{j}}-|\mathbf{r}_{\mathrm{ij}}|$,
where $\mathbf{r}_{\mathrm{ij}}=\mathbf{r}_{\mathrm{i}}-\mathbf{r}_{\mathrm{j}}$ and $\mathbf{n}=\frac{\mathbf{r}_{\mathrm{ij}}}{|\mathbf{r}_{\mathrm{ij}}|}$.
The second term in the normal interation is the viscous damping term,
here $\mathbf{v}_{\mathrm{n,rel}}=(\mathbf{v}_{\mathrm{j}}-\mathbf{v}_{\mathrm{i}})\cdot\mathbf{n}\ \mathbf{n}$,
and $\eta_{\mathrm{n}}=-2\sqrt{\frac{5}{6}}\frac{\log(e)}{\sqrt{\pi^{2}+(\log(e))^{2}}}\sqrt{\frac{3}{2}k_{\mathrm{nd}}m_{\mathrm{eff}}}$
, where $k_{\mathrm{nd}}=\frac{4}{3}E_{\mathrm{eff}}a$ with $a=\sqrt{\delta_{\mathrm{ij}}R}$
,$m_{\mathrm{eff}}=\frac{M_{\mathrm{i}}M_{\mathrm{j}}}{M_{\mathrm{i}}+M_{\mathrm{j}}}$ and $e$ is the coefficient of restitution.

In addition to this, there is a tangential component, 
\begin{equation}
\mathbf{F}_{\mathrm{te}} + \mathbf{F}_{\mathrm{t,damp}} = 
\begin{cases} 
  \int_{\mathrm{t_0}}^t k_t a \mathbf{v}_{\mathrm{t,rel}}(\tau)\text{d}\tau - \eta_n \mathbf{v}_{\mathrm{t,rel}} \\[1ex] 
  \qquad \text{if } \mathbf{F}_{\mathrm{te}} + \mathbf{F}_{\mathrm{t,damp}} < \mu_t |\mathbf{F}_n| \\[2ex]
  \mu_t |\mathbf{F}_n| \\[1ex] 
  \qquad \text{if } \mathbf{F}_{\mathrm{te}} + \mathbf{F}_{\mathrm{t,damp}} > \mu_t |\mathbf{F}_n| 
\end{cases}
\end{equation}
where $\mathbf{v}_{\mathrm{t,rel}}=\mathbf{v}_{\mathrm{t}}-(R_{\mathrm{i}}\boldsymbol{\Omega}_{\mathrm{i}}+R_{\mathrm{j}}\boldsymbol{\Omega}_{\mathrm{j}})\times\mathbf{n}$,
$\mathbf{v}_{\mathrm{t}}=\mathbf{v}_{\mathrm{r}}-\mathbf{v}_{\mathrm{r}}\cdot\mathbf{n}\ \mathbf{n}$,
$\mathbf{v}_{\mathrm{r}}=\mathbf{v}_{\mathrm{j}}-\mathbf{v}_{\mathrm{i}}$, $\mu_{\mathrm{t}}$ is the
friction co-efficient, $k_{\mathrm{t}}=8\left(\frac{2-\nu_{\mathrm{i}}}{G_{\mathrm{i}}}+\frac{2-\nu_{\mathrm{j}}}{G_{\mathrm{j}}}\right)^{-1}$
with $G_{\mathrm{i}}=E_{\mathrm{i}}/(2(1+\nu_{\mathrm{i}}))$. 
These interactions are implemented in LAMMPS as \texttt{pair\_style granular} with \texttt{hertz/material} for normal interaction and \texttt{tangential mindlin} for tangential interaction. 
\subsection{Fluid-Structure Interaction}
\label{FSIdetails}
In this section we describe how particle motion influences the populations $f_i$'s and how forces acting on the particles are calculated from $f_i$'s.

\subsubsection{How $f_i$'s are modified due to particle motion}
A given lattice node with $d<1$ has a velocity $\mathbf{u}_p$ defined by the
particle's translational and rotational velocity.
In addition to the standard BGK collision Eq. \ref{eq:LBMupdate}, we have to apply the Brinkman penalization as an external force, which is applied to the system through the exact difference scheme of Kupershtokh \citep{trunkRevisitingHomogenizedLattice2021}.
First we calculate 
\begin{equation}
    \mathbf{u}^+ = \mathbf{v} + (1-d)[\mathbf{u}_p-\mathbf{v}]
\end{equation} and  
then the populations are modified by 
\begin{equation}
    S_i = f_i^{eq}(\rho ,\mathbf{u}^+)-f_i^{eq}(\rho,\mathbf{v}) 
\end{equation}
with $S_i$'s added to the RHS of Eq. \ref{eq:LBMupdate} before streaming. The physical
velocity is calculated using
\begin{equation}
    \mathbf{u} ^* = \mathbf{v} + \frac{1}{2}(1-d)[\mathbf{u}_p-\mathbf{v}]
    \label{eq:physicalVelocity}
\end{equation} as described in \citep{krugerLatticeBoltzmannMethod2017a}.

An external body force $\mathbf{F}$ acting on the fluid is applied by the very same
construction and simply adds to the shifted velocity,
\begin{equation}
    \mathbf{u}^+ = \mathbf{v} + \frac{\Delta t}{\rho}\mathbf{F} + (1-d)[\mathbf{u}_p-\mathbf{v}],
    \label{eq:forcedShift}
\end{equation}
so that the body force and the penalization are realized by a single exact
difference step, with the corresponding term $\Delta t\,\mathbf{F}/(2\rho)$ added to
Eq. \ref{eq:physicalVelocity}. In the hindered settling simulations, $\mathbf{F}$ is
the compensating force of Sec. \ref{sec:hinderedSettling}, which keeps the mean
velocity of the fluid at zero.

Comparing this with the penalized momentum equation Eq. \ref{eq:penalized}, the
scheme relaxes the local velocity towards $\mathbf{u}_{\mathrm{p}}$ by the
fraction $(1-d)$ in every timestep, i.e. the discrete penalization parameter is
$\eta_{\mathrm{p}}=\Delta t/(1-d)$ and the corresponding local permeability is
$K=\nu\eta_{\mathrm{p}}$. In the fully fluid limit $d=1$ the forcing vanishes and
the scheme reduces to standard BGK, while inside a body $d=0$ and the penalization
is at its strongest. The attainable penalization is thus bounded by the timestep
rather than chosen freely, so that at a fixed resolution the residual slip, and
hence the position of the effective hydrodynamic surface, retains a weak dependence
on $\tau$. This is the origin of the $\tau$ dependence reported in Appendix
\ref{DragForceLadd} and Appendix \ref{cubetermianlVelocity}.
\subsubsection{How forces on particles are calculated from $f_i$'s}
We use the Galilean invariant momentum exchange formulation from \citep{wen2014galilean}, where the hydrodynamic force acting on a given particle is calculated by
\begin{equation}
    \mathbf{F}^{\mathrm{H}}(t) = -\sum_{\mathbf{x}\in B} \sum_{\mathrm{i=1}}^q \left[(\mathbf{C}_{\mathrm{i}} - \mathbf{u}_{\mathrm{p}})f_{\mathrm{i}}(\mathbf{x}+\mathbf{C}_{\mathrm{i}},t)+(\mathbf{C}_{\mathrm{i}} + \mathbf{u}_{\mathrm{p}})f_{\mathrm{\tilde{i}}}(\mathbf{x},t)\right]
    \label{eq:momentumExchange}
\end{equation}
where $\tilde{i}$ is the direction opposite to $i$. The sum is restricted to the
lattice links that cross the fluid-solid interface, i.e. to the nodes $\mathbf{x}$
with $d<1$ that belong to the particle and whose neighbour $\mathbf{x}+\mathbf{C}_{\mathrm{i}}$
is a pure fluid node with $d=1$. Nodes in the interior of a particle, which have no
fluid neighbour, do not contribute. The corresponding torque
about the particle pivot $\mathbf{x}_{\mathrm{p}}$ follows as
\begin{equation}
    \mathbf{T}^{\mathrm{H}}(t) = \sum_{\mathbf{x}\in B} (\mathbf{x}-\mathbf{x}_{\mathrm{p}}) \times \mathbf{F}^{\mathrm{H}}(\mathbf{x},t),
    \label{eq:momentumExchangeTorque}
\end{equation}
where $\mathbf{F}^{\mathrm{H}}(\mathbf{x},t)$ is the per-node contribution to Eq.
\ref{eq:momentumExchange}. For the periodic directions of the domain, the lever arm
$\mathbf{x}-\mathbf{x}_{\mathrm{p}}$ is taken with respect to the nearest periodic
image of the pivot, i.e. each component is shifted by the corresponding domain length
whenever it exceeds half of it. Without this minimum image convention, a particle
straddling a periodic boundary would be assigned lever arms of nearly the domain
length for the nodes on the far side.

\subsubsection{Representation of a particle in the fluid}
We initialize a reference lattice, whose dimensions bound the object we would like to
represent, with a porosity field representing the
\texttt{.stl} file using \texttt{STLreader} capability of OpenLB. The reference
lattice is resolved at half the spacing of the fluid lattice and its nodes are set
to $0$ or $1$ depending on whether they fall inside or outside the surface mesh, so
that the reference porosity is an exact voxelization of the given geometry. The same
voxelization provides the volume, the mass and the inertia tensor of the body, which
are passed to LAMMPS through its \texttt{infile} feature, so that the fluid and the
DEM side use a consistent description of the rigid body.

Given a pivot to
place the particle (taken to be its center of mass) and the orientation about the pivot, we can map
a given lattice node in our domain to a point in the
reference lattice and therefore use the predefined porosity field to embed the
particle in our lattice. The porosity assigned to a node of the fluid lattice is the
average of the reference porosity over the mapped node and its $q-1$ neighbours in
the reference lattice. This smooths the voxelized interface over the width of one
reference cell, so that the surface of the body is located approximately at the
$d=1/2$ level set. The mapping uses the same minimum image convention as above, so
that a particle whose support crosses a periodic boundary is embedded as one body.

The solid velocity entering the forcing follows from the rigid body kinematics of the
body $k$ covering the node,
\begin{equation}
    \mathbf{u}_{\mathrm{p}}(\mathbf{x},t) = \mathbf{v}_{\mathrm{k}}(t) + \boldsymbol{\omega}_{\mathrm{k}}(t)\times\left(\mathbf{x}-\mathbf{x}_{\mathrm{k}}(t)\right),
    \label{eq:solidVelocity}
\end{equation}
for center of mass $\mathbf{x}_{\mathrm{k}}$, translational velocity $\mathbf{v}_{\mathrm{k}}$
and angular velocity $\boldsymbol{\omega}_{\mathrm{k}}$.

During the motion of particles in the simulation, this
procedure is repeated to keep track of the porosity field. Updating the porosity
field after a displacement or a rotation of a body does not require a re-evaluation
of all the nodes of the domain. Between two timesteps the surface of a body may not
travel further than the distance between adjacent nodes, so that only the nodes
currently covered by a body and a surrounding growth layer have to be revisited.
The cost of the geometry update, like that of the force evaluation, is therefore
proportional to the surface and not to the volume of the bodies. The porosity
embedding and the surface force integration are implemented as platform-transparent
operators in OpenLB and are executed on the GPU as described in
\citep{kummerlaenderHFF2026,kummerlaenderCMAME2026}.


\section{Drag force on cubic array of spheres}
\label{DragForceLadd}
\begin{figure}
    \centering
    \includegraphics[width=\linewidth]{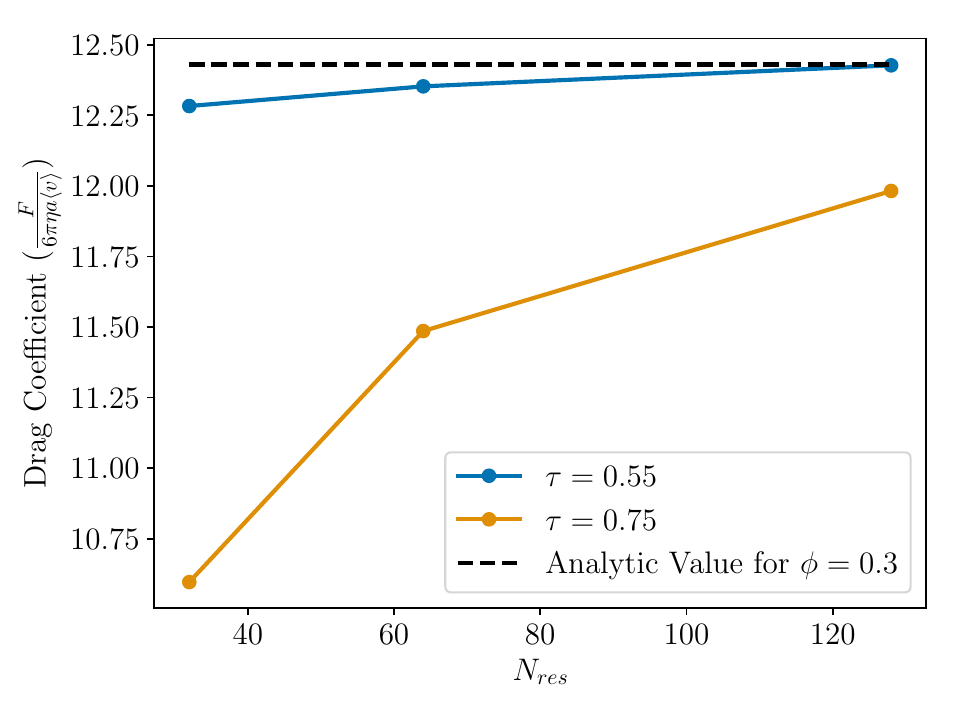}
    \caption{Drag force for different $\tau$ and $N_{\mathrm{res}}$ at $\phi=0.3$. }
    \label{fig:laddDragValid}
\end{figure}
In this section we describe the validation associated with 
calculating the drag force on a cubic array of spheres. For 
a packing fraction $\phi=0.3$ we run the simulation for 
different values of $\tau$ and $N_{\mathrm{res}}$. The results are 
shown in Fig. \ref{fig:laddDragValid}, where we observe a 
convergence to thee analytical value. Since the force acting 
on the sphere at steady state is equal to the applied force to 
the fluid for all values of $\tau$, the average velocity is 
the value that changes as we change the parameter. Therefore, 
we observe a higher velocity than the analytical prediction 
for higher values of $\tau$, which we observe also in the 
single sphere settling.
\section{Sphere settling velocity}
\label{spheretermianlVelocity}
\begin{figure}[t]
\includegraphics[width=1\linewidth]{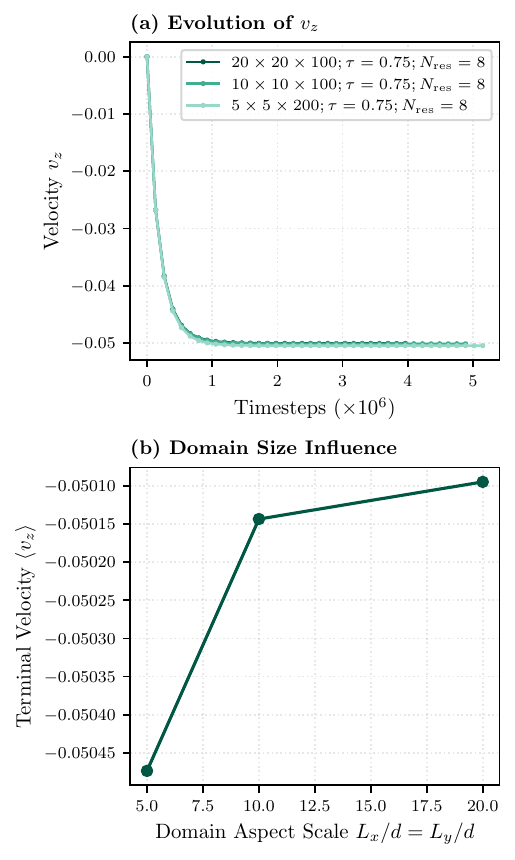}
\caption{For $N_{\mathrm{res}}=8,$ i.e the diameter is resolved upto 8 l.u's. }\label{fig:domain}
\end{figure}
\begin{figure}[t]
\includegraphics[width=1\linewidth]{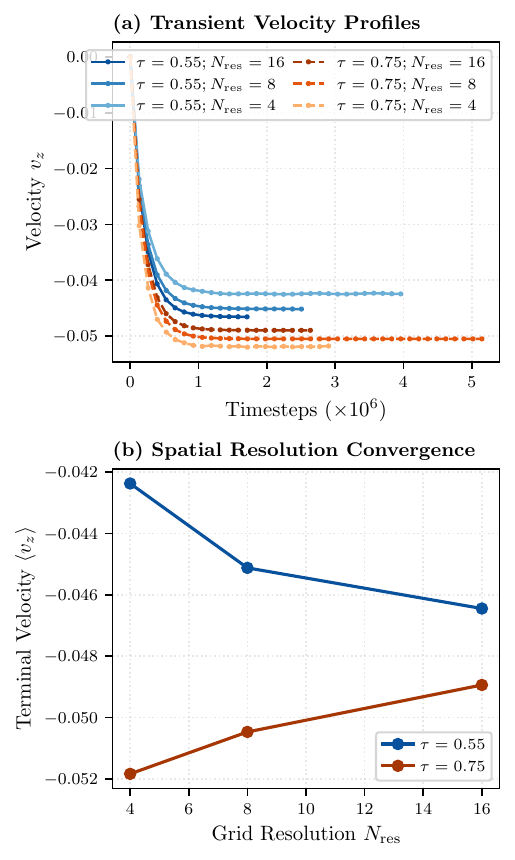}
\caption{Settling velocity of sphere as one increases the resolution with which a diameter
is resolved for two different $\tau$ values. }\label{fig:res_vs_vz}
\end{figure}
\begin{figure}[t]
    \centering
    \includegraphics[width=1\linewidth]{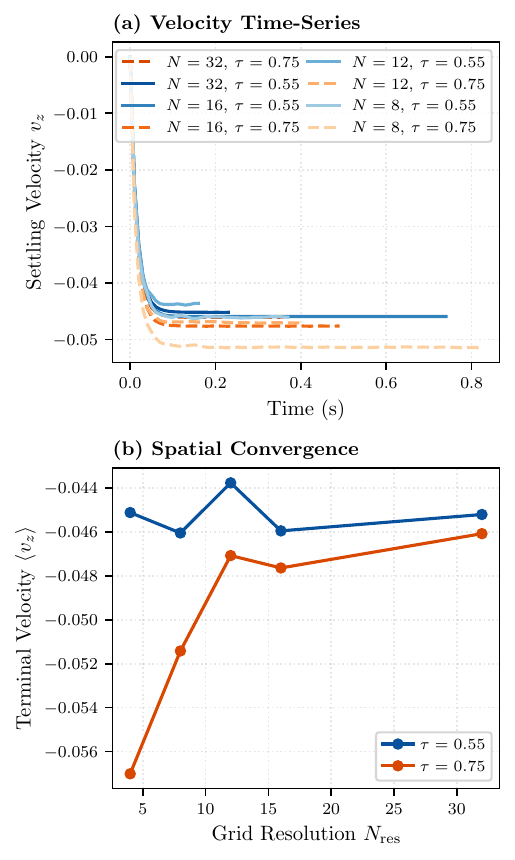}
    \caption{Terminal velocities of a single cube for different resolutions $N_{\mathrm{res}}$ and for two different relaxation times $\tau$.}
    \label{fig:cubeTerminal}
\end{figure}

\begin{figure}[h]
    \centering
    \includegraphics[width=\linewidth]{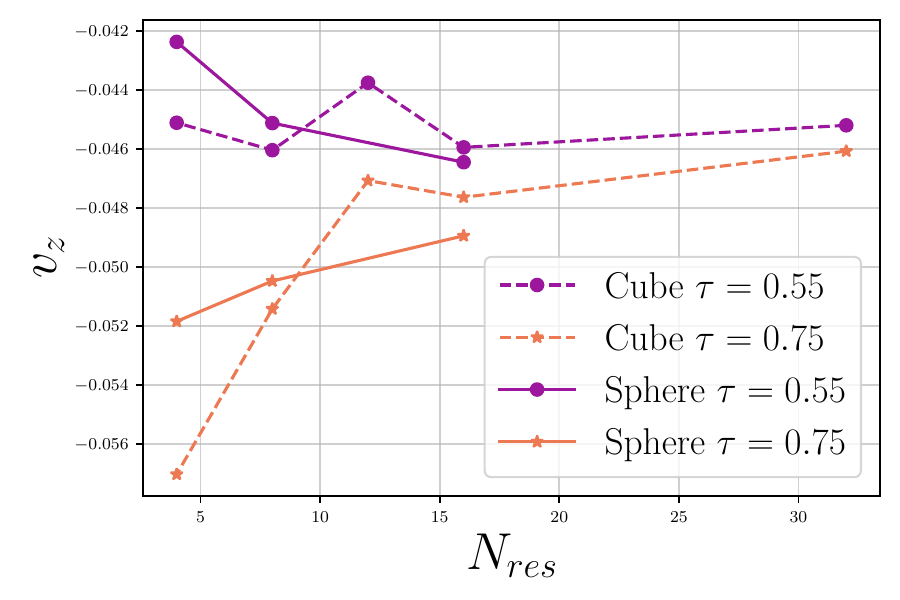}
    \caption{Comparison of cube and sphere settling velocity for different resolutions and relaxation times.}
    \label{fig:cubeSpheres}
\end{figure}
In this section we describe sphere settling simulations. We will calibrate the system by comparing the terminal velocity of
the sphere with a known ground truth. The system under study has $d=0.35\times10^{-3}$
m diameter, $\rho_{\mathrm{p}}=2500$ kg/m$^{3}$, $\rho_{\mathrm{f}}=1000$ kg/m$^{3}$
and $\mu=10^{-3}$ Pa.s (same as the one used in \citet{rettingerCoupledLatticeBoltzmann2017}).
The terminal velocity of a single particle
is taken to be $\approx 0.048$ m/s.
We need to study the influence of $\tau$, $N_{\mathrm{res}}$, on the numerical
terminal velocity. First, we study
different domain sizes as multiples of the diameter of the sphere to ascertain the dependence of domain size on the terminal velocity. For the simulation,  
we use bounceback boundary condition for the top and bottom walls, 
a pressure boundary condition for the side walls as implemented
in \texttt{interpolatedPressureBoundary} in OpenLB. 
As shown in Fig. \ref{fig:domain}, the effect of the domain walls only changes the terminal velocity by less than 1$\%$. Therefore for further studies we will use the 
$5\times 5 \times 200$ box size.

In Fig. \ref{fig:res_vs_vz} we plot the variation of $v_{\mathrm{z}}$ as
we increase the resolution of the sphere diameter for two relaxation times
$\tau$. As seen in the graph, as we increase the resolution of the
system, the value seems to converge to a value around the expected
value of around $\sim 0.048$. As resolution is increased, the effect 
of relaxation time $\tau$ reduces, indicating that the simulation is 
converging. 

\section{Cube Settling Velocity}
\label{cubetermianlVelocity}

Similar to the preceding sphere study, we study the effect of $\tau$ 
and $N_{\mathrm{res}}$ on the cube settling velocity. In Fig. 
\ref{fig:cubeTerminal}, we show the dependence of the cube velocity on 
different parameters. Again, we observe a convergence to a value close 
to $\sim -0.046$ m/s. This is close to, but faster than what is 
predicted from empirical relations - Haider's formula \cite{haiderDragCoefficientTerminal1989} predicts $\sim 0.0437$ m/s. 
From existing literature \cite{marquardtNovelParticleDecomposition2024,kunduSettlingDynamicsNonBrownian2025,seyed-ahmadiSedimentationInertialMonodisperse2021} we 
expect the cube to settle slower than the sphere. However, we observe 
that the sphere and the cube terminal velocities are close to each 
other as shown in Fig. \ref{fig:cubeSpheres}.

\section{Comparison with \citet{seyed-ahmadiSedimentationInertialMonodisperse2021}}
\label{comparisonAhmadiSection}
\begin{figure}
    \centering
    \includegraphics[width=\linewidth]{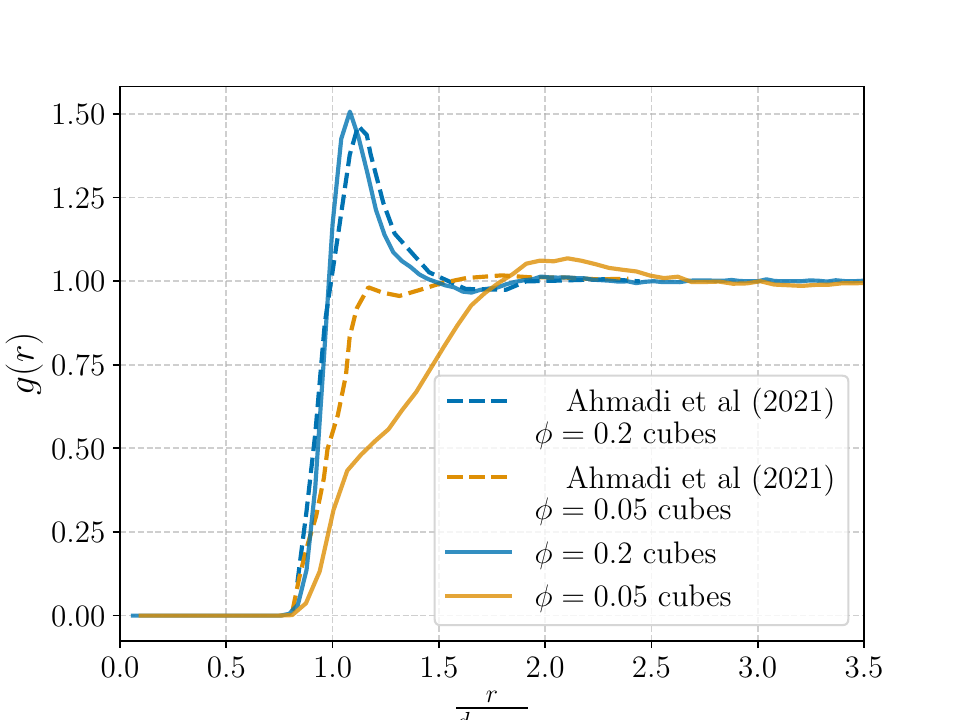}
    \caption{Comparison with \citet{seyed-ahmadiSedimentationInertialMonodisperse2021}}
    \label{fig:comparisonAhmadi}
\end{figure}
We compare the radial distribution function obtained for cubes 
from \citet{seyed-ahmadiSedimentationInertialMonodisperse2021} 
at two different packing fractions as shown in Fig. \ref{fig:comparisonAhmadi}. Although, their simulation is at a 
higher $Re$ than ours, we observe that at higher packing 
fraction, the local structure is very similar. However, at 
lower packing fractions, there is more clustering for the 
higher $Re$ case. We also note the that RDF for $\phi=0.05$ 
from \citet{seyed-ahmadiSedimentationInertialMonodisperse2021} 
and RDF for $\phi=0.1$ from Fig. \ref{fig:corr_v} for cubes is quantitatively 
very similar. 
\section{Lubrication force between the particles}
\label{LubricationForce}

\begin{figure}[h]
    \centering
    \includegraphics[width=\linewidth]{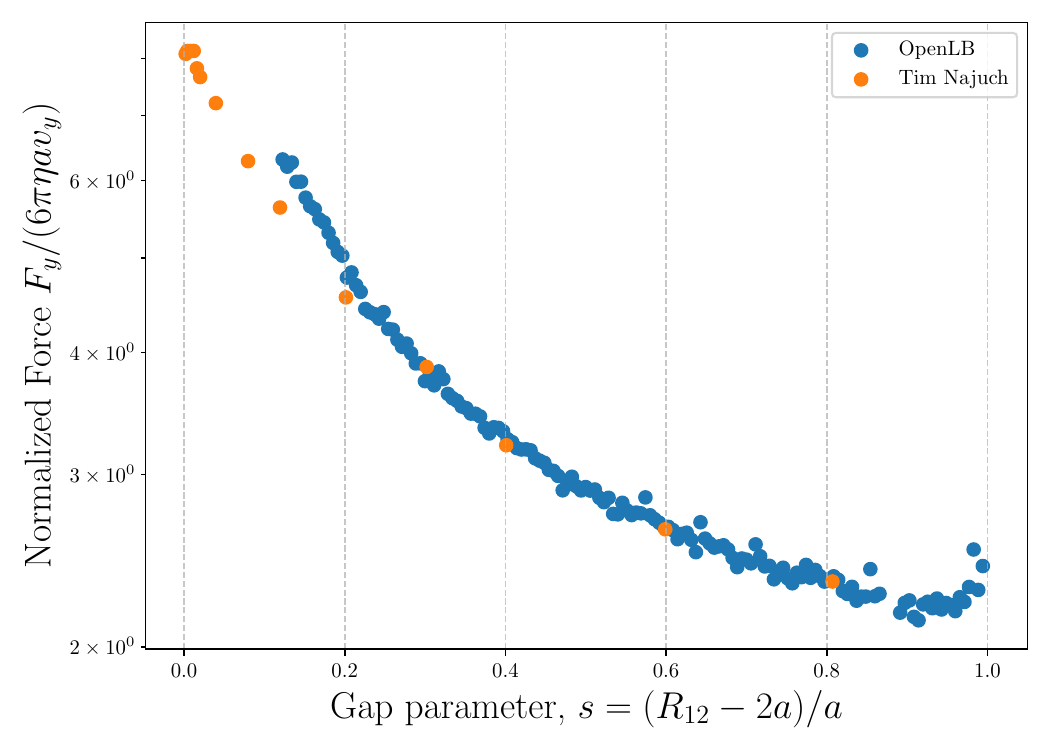}
    \caption{The force acting on a sphere compared with its stokes drag for two sphere approaching each other along the y direction.}
    \label{fig:lubrication}
\end{figure}
As discussed in 
\citet{laddNumericalSimulationsParticulate1994,rettingerCoupledLatticeBoltzmann2017,najuchSimulationDenseSuspensions,tencateParticleImagingVelocimetry2002} there is a 
repulsive force that acts between two spheres as they come 
close to each other. This is due to the fluid getting squeezed 
out of the gap between the two spheres. Numerical simulation by 
\citet{najuchSimulationDenseSuspensions} show that the 
partially saturated method underestimates the repulsion between 
the two spheres approaching each other, necessitating an 
addition of a lubricating force. We find the same result here 
for this system Fig. \ref{fig:lubrication}. However, we have 
not included a lubrication force in the simulations performed. 

\section{Hindered Settling of Spheres and Cubes}
\begin{figure}[h]
    \centering
    \includegraphics[width=\linewidth]{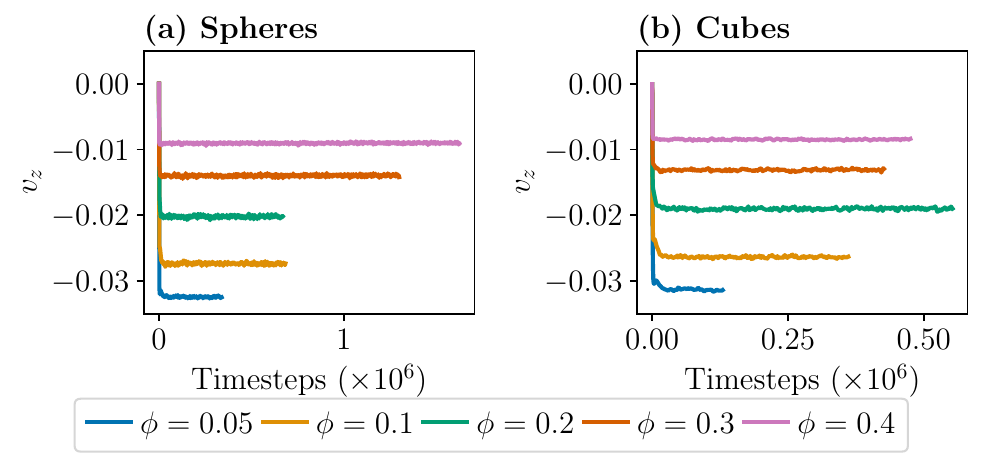}
    \caption{Vertical velocity component of sphere and cubes at different packing fractions.}
    \label{fig:termVcubesSpheres}
\end{figure}
In this section, we present the evolution of settling velocity 
for bulk suspension at different packing fraction. The results 
correspond to Fig. \ref{fig:hinderedSettlingCubesSpheres}. We 
observe that cube always settles slower than than sphere at 
all packing fractions. However, when compared to \cite{marquardtNovelModelDirect2024a} reports a $25\%$ difference between the velocities of cubes and spheres, while in our system the difference settling velocities is $\sim 6\%$.
\clearpage
\newpage

\section{Validation of clump representation}
\label{validclumosection}
\begin{figure}
    \centering
    \includegraphics[width=0.8\linewidth]{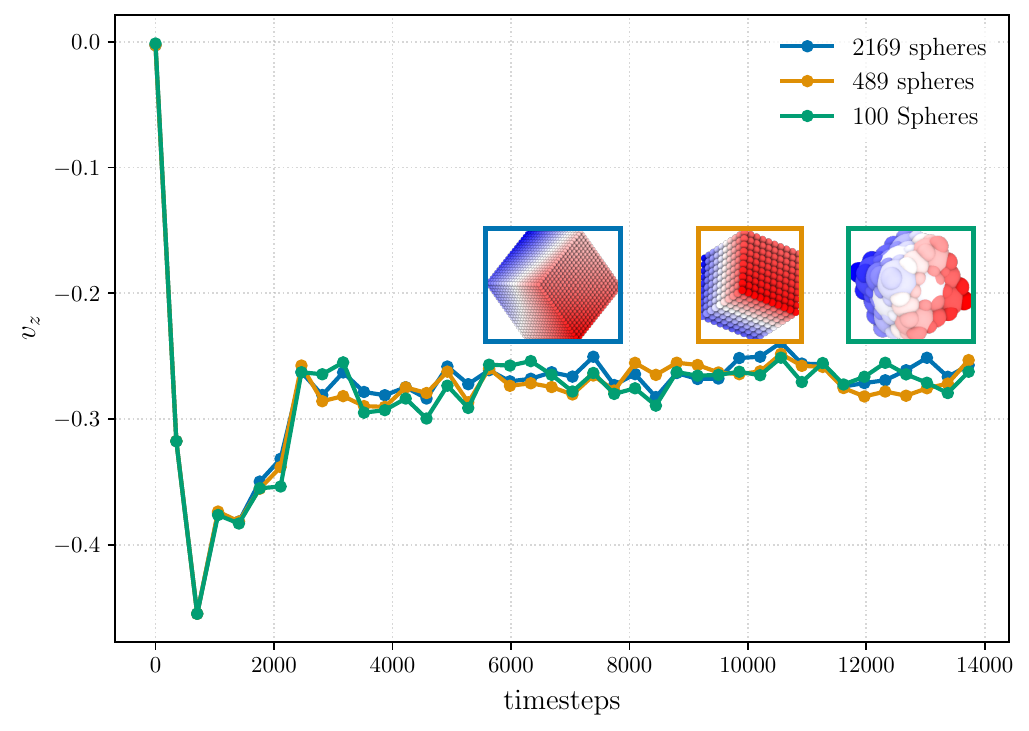}
    \caption{Evolution of $v_{\mathrm{z}}$ for hindered settling with two different ``clump'' representation of cube with $489$ spheres and $100$ spheres respectively. The corresponding clumps are shown in the inset.}
    \label{fig:validClump}
\end{figure}
In this section we present a validation test for clump 
representation. We perform a hindered settling of 125 cubes. 
The two representation are shown in the insets. The 100 sphere 
representation is created using the 
\citet{angelidakisCLUMPCodeLibrary2021} library while the 489 and 2169 
sphere representations are created by placing spheres in a grid 
on the cube surfaces. As shown in Fig. \ref{fig:validClump}, 
$v_z$ evolves similarly for both systems. 
\newpage

\newpage


\newpage

\end{document}